\documentclass[12pt,preprint]{aastex}
\usepackage{bm}
\usepackage{threeparttable}
\usepackage{amsmath,amssymb}
\usepackage{textcomp}
\usepackage{booktabs}
\usepackage{lscape}
\makeatletter
\newcommand{\tblcaption}[1]{\def\@captype{table}\caption{#1}}
\makeatother
\usepackage{subfigure}

\begin{document}


\title{OGLE-2012-BLG-0950Lb: THE FIRST PLANET MASS MEASUREMENT FROM ONLY MICROLENS PARALLAX AND LENS FLUX}


\author{N. Koshimoto\altaffilmark{1,m}, A. Udalski\altaffilmark{2,o}, J.P. Beaulieu\altaffilmark{3,p}, T.Sumi\altaffilmark{1,m}, D.P. Bennett\altaffilmark{4,5,m},
I.A. Bond\altaffilmark{6,m}, N. Rattenbury\altaffilmark{7,m}, A. Fukui\altaffilmark{8,m}, V. Batista\altaffilmark{3,p}, J.B. Marquette\altaffilmark{3,p}, S.Brillant\altaffilmark{9,p}}

\and

\author{F. Abe\altaffilmark{10}, Y. Asakura\altaffilmark{10}, A. Bhattacharya\altaffilmark{4,5}, M. Donachie\altaffilmark{7}, M. Freeman\altaffilmark{7}, Y. Hirao\altaffilmark{1},
Y. Itow\altaffilmark{10}, M.C.A. Li\altaffilmark{7}, C.H. Ling\altaffilmark{6}, 
K. Masuda\altaffilmark{10}, Y. Matsubara\altaffilmark{10}, T. Matsuo\altaffilmark{1}, Y. Muraki\altaffilmark{10}, M. Nagakane\altaffilmark{1}, 
K. Ohnishi\altaffilmark{11}, H. Oyokawa\altaffilmark{10}, To. Saito\altaffilmark{12}, 
A. Sharan\altaffilmark{7}, H. Shibai\altaffilmark{1}, D.J. Sullivan\altaffilmark{13}, D. Suzuki\altaffilmark{4,5}, P.J. Tristram\altaffilmark{14}, A. Yonehara\altaffilmark{15}\\
(MOA Collaboration)}
\author{S. Koz{\l}owski\altaffilmark{2}, P. Pietrukowicz\altaffilmark{2}, 
R. Poleski\altaffilmark{2,16}, J. Skowron\altaffilmark{2},
I. Soszy{\'n}ski\altaffilmark{2}, M.\,K. Szyma{\'n}ski\altaffilmark{2}, 
K. Ulaczyk\altaffilmark{2}, {\L}. Wyrzykowski\altaffilmark{2}\\
(OGLE Collaboration)}

\altaffiltext{1}{Department of Earth and Space Science, Graduate School of Science, Osaka University, 1-1 Machikaneyama, Toyonaka, Osaka 560-0043, Japan}
\altaffiltext{2}{Warsaw University Observatory, Al. Ujazdowskie 4, 00-478 Warszawa, Poland}
\altaffiltext{3}{Sorbonne Universites, CNRS, UPMC Univ Paris 06, UMR 7095, Institut d'Astrophysique de Paris, F-75014, Paris, France}
\altaffiltext{4}{Department of Physics, University of Notre Dame, Notre Dame, IN 46556, USA}
\altaffiltext{5}{Laboratory for Exoplanets and Stellar Astrophysics, NASA/Goddard Space Flight Center, Greenbelt, MD 20771, USA}
\altaffiltext{6}{Institute of Information and Mathematical Sciences, Massey University, Private Bag 102-904, North Shore Mail Centre, Auckland, New Zealand}
\altaffiltext{7}{Department of Physics, University of Auckland, Private Bag 92019, Auckland, New Zealand}
\altaffiltext{8}{Okayama Astrophysical Observatory, National Astronomical Observatory, 3037-5 Honjo, Kamogata, Asakuchi, Okayama 719-0232, Japan}
\altaffiltext{9}{European Southern Observatory (ESO), Karl-Schwarzschildst. 2, D-85748 Garching, Germany}
\altaffiltext{10}{Institute for Space-Earth Environmental Research, Nagoya University, Nagoya 464-8601, Japan}
\altaffiltext{11}{Nagano National College of Technology, Nagano 381-8550, Japan}
\altaffiltext{12}{Tokyo Metropolitan College of Industrial Technology, Tokyo 116-8523, Japan}
\altaffiltext{13}{School of Chemical and Physical Sciences, Victoria University, Wellington, New Zealand}
\altaffiltext{14}{Mt. John Observatory, P.O. Box 56, Lake Tekapo 8770, New Zealand}
\altaffiltext{15}{Department of Physics, Faculty of Science, Kyoto Sangyo University, Kyoto 603-8555, Japan}
\altaffiltext{16}{Department of Astronomy, The Ohio State University, 140 West 18th Avenue, Columbus, OH 43210, USA}
\altaffiltext{m}{Microlensing Observations in Astrophysics (MOA) Collaboration}
\altaffiltext{o}{Optical Gravitational Lensing Experiment (OGLE) Collaboration}
\altaffiltext{p}{Probing Lensing Anomalies NETwork (PLANET) Collaboration}


\begin{abstract}
We report the discovery of a microlensing planet OGLE-2012-BLG-0950Lb with the planet/host mass ratio of $q \simeq 2 \times 10^{-4}$.
A long term distortion detected in both MOA and OGLE light curve can be explained by the microlens parallax
due to the Earth's orbital motion around the Sun. 
Although the finite source effect is not detected, we obtain the lens flux by the high resolution Keck AO observation. 
Combining the microlens parallax and the lens flux reveal the nature of the lens: 
a planet with mass of $M_{\rm p} = 35^{+17}_{-9} M_{\oplus}$ is orbiting around a M-dwarf with mass of $M_{\rm host} = 0.56^{+0.12}_{-0.16} M_{\odot}$
with a planet-host projected separation of $r_{\perp} =2.7^{+0.6}_{-0.7}$ AU located at $D_{\rm L} = 3.0^{+0.8}_{-1.1}$ kpc from us. 
This is the first mass measurement from only microlens parallax and the lens flux without the finite source effect.
In the coming space observation-era with {\it Spitzer}, {\it K2}, {\it Euclid}, and {\it WFIRST},
we expect many such events for which we will not be able to measure any finite source effect.
This work demonstrates an ability of mass measurements in such events.
\end{abstract}


\section{Introduction}
Gravitational microlensing is a technique by which planets can be detected without measurements of light from
the host star \citep{mao91,gouloe92,gau12}.
Microlensing can detect planets that are difficult to detect by other methods such as planetary systems in the Galactic Bulge (e.g., Batista et al. 2014), 
planets around late M-dwarfs or brown dwarfs \citep{ben08, sum16}, and even free floating planets not hosted by any stars \citep{sum11}.
Compared to other techniques, microlensing is sensitive to Earth mass planets \citep{ben96} orbiting just outside of the snow line where 
the core accretion theory \citep{ida04} predicts is the most active planet formation region.
Microlensing observations so far have revealed a population of planets beyond the snow line \citep{gou10, sum10, cas12, shv16, suz16}.
\citet{suz16} finds a steeper slope with $dN/d\log {q} \sim  q^{-0.9}$ and 
a break (and possible peak) in the mass ratio function at $q \sim 1.0 \times 10^{-4}$.
We are capable of studying the distance distribution of planets in our Galaxy via microlensing. \citet{pen16} suggests the possibility of a lack of planets 
in the Galactic bulge.
The detection of extra solar planets by gravitational microlensing presents a number of challenges.

Firstly, gravitational microlensing is an extremely rare phenomenon with a probability of one per one million stars and 
a planetary deviation lasts for only hours or a few days.
For these reasons, microlensing observations for exoplanets are conducted towards the Galactic bulge, the most crowded field in our Galaxy.
Whereas hundreds of planets are detected by the radial velocity (RV) method \citep{but06,bon13} and thousands of planetary candidates are detected by
the {\it Kepler} telescope \citep{bor10} to date, the microlensing method has been used to detect about 50 exoplanets so far.

Several survey groups have been conducting high cadence survey observations using their telescopes with 
wide FOV cameras in different time zones.
The Microlensing Observations in Astrophysics (MOA; Bond et al. 2001, Sumi et al. 2003) group uses the 2.2-deg$^2$ FOV MOA-cam3 \citep{sak08} CCD camera mounted on 
the 1.8 m MOA-II telescope at the Mt.\ John University Observatory in New Zealand and alerts the community about 600 microlensing events per year.
The Optical Gravitational Lensing Experiment group (OGLE; Udalski 2003) upgraded their camera to the 1.4-deg$^2$ FOV OGLE-IV camera in 2010 \citep{uda15a}
and discovered more than 2000 microlensing events per year in the last few years with the camera mounted on the 1.3 m Warsaw telescope at the Las Campanas Observatory, Chile.
The Wise observatory group in Israel also conducts microlensing observations \citep{gor10}.
In 2015, the Korean Microlensing Telescope Network (KMTNet; Kim et al. 2016) also started their survey observations.
Now the equipment requirements for second-generation microlensing surveys \citep{gau09,gau12} are 
fulfilled and the number of planet detections is increasing over the next few years.

 Measuring the mass of a lens $M_{\rm L}$ and the distance to the lens system $D_{\rm L}$ is challenging.
There are three observables in microlensing which can yield a mass-distance relation of the lens system:
the angular Einstein radius $\theta_{\rm E}$, the microlens parallax $\pi_{\rm E}$  and the lens flux.
The first two of these yield each mass-distance relation by combining the following relationship between them;
\begin{equation}
M_{\rm L} = \frac{\theta_{\rm E}}{\kappa \pi_{\rm E}}
\end{equation}
with the definitions of $\pi_{\rm E}$, $\pi_{\rm E} \equiv \pi_{\rm rel} / \theta_{\rm E}$, where $\kappa$ is a constant and 
$\pi_{\rm rel} \equiv {\rm AU} (1/D_{\rm L} - 1/D_{\rm S})$.
One can calculate the mass and distance of the lens system if we can measure any two of these quantities.
We can measure $\theta_{\rm E}$ in the following manner. Included in most models explaining planetary microlensing 
light curve data is the source star radius in units of $\theta_{\rm E}$: $\rho \equiv \theta_*/\theta_{\rm E}$.
By estimating the angular radius of the source star, $\theta_*$, by an analysis of the source star's color and magnitude,
and using our modeled value of $\rho$, we arrive at an estimate of $\theta_{\rm E}$.
Microlens parallax can be observed only in relatively rare events and lens flux measurements need follow-up observations
with high resolution imaging by an 8-m class telescope.
Therefore only half of planetary events published so far are detected with lens mass measurements and masses of the other half planetary systems are just given
their probability distributions by a Bayesian analysis (e.g., Beaulieu et al. 2006; Bennett et al.2014; Koshimoto et al.2014; Skowron et al 2015).

In the microlensing planetary events published so far, there are events with the mass measurements from the angular Einstein radius and microlens parallax 
(e.g., Bennett et al. 2008; Gaudi et al. 2008; Muraki et al. 2011), from the angular Einstein radius and the lens flux (e.g., Bennett et al. 2006; Batista et al. 2015; Bennett et al. 2015),
and from all three relations (e.g., Dong et al. 2009; Bennett et al. 2010; Beaulieu et al. 2016; Bennett et al. 2016), 
but events with mass measurement from only microlens parallax and the lens flux have not been published to date.
This is simply because the angular Einstein radius is observed much commonly than microlens parallax as mentioned above.
However, it has been possible to measure precise microlens parallax by observing simultaneously from space and ground thanks to 
the {\it Spitzer} microlensing campaign \citep{cal15,uda15b,yeeet15,zhu15}.
Also, {\it K2} campaign 9 ({\it K2}C9), started in April 2016, has surveyed the Galactic bulge for three months to date.
By combining {\it K2}C9 data and ground-based survey data, it is expected to measure microlens parallax for more than 120 events \citep{hen15}. 
These next generation space- and ground-based simultaneous observations for microlensing can measure microlens parallax almost regardless of the event timescale.
Microlens parallax should become a more common observable rather than the angular Einstein radius in coming next generation, 
thus the mass measurement without the angular Einstein radius should be important \citep{yee15}.

This paper reports an analysis of a microlensing planetary event OGLE-2012-BLG-0950, which is the first event where a mass measurement is possible from 
only the measurements of the microlens parallax and lens flux.
The survey observations of this event are described in Section \ref{sec-obs}. 
Section \ref{sec-reduction} explains our data reduction procedure.
Section \ref{sec-model} shows our modeling results. 
We show the constraint on the angular Einstein radius by the source angular radius derived from the color and light curve modeling in Section \ref{sec-color}.
 In Section \ref{sec-KECK} we describe our Keck observations, the constraints on the excess flux and calculate the probability of the contamination to the  excess flux.
In Section \ref{sec-lens}, we derive the lens properties by combining microlens parallax and the lens flux. 
Finally, Section \ref{sec-disc} discusses and concludes the results of this work.

\section{Observations}\label{sec-obs}

Microlensing event OGLE-2012-BLG-0950 was discovered and alerted by the OGLE Early Warning System (EWS) on 21 June 2012 (HJD$' \equiv $ HJD - 2450000 $\sim $ 6100) as 
a new event located at $(R.A., Dec.)_{J2000}$ = (18:08:04.62, -29:43:53.7) or $(l, b) = (1.765^{\circ},-4.634^{\circ})$. 
Another survey group, MOA, independently found the event and alerted that as MOA-2012-BLG-527 on 9 August 2012.
The observations by OGLE were conducted on the $I$-band and $V$-band and 
the observations by MOA were conducted by the custom MOA-Red filter which is similar to the sum of the standard Cousins $R$- and $I$-band filters.
MOA also observed the event in the $I$-band using the B$\&$C telescope, a 61 cm telescope for follow-up observation at the same site.
The observed light curve is shown in Figure \ref{fig-models}.

The anomaly part of this event appeared as a small dip around HJD$' \sim $ 6149 mainly in the MOA data.
The MOA-II telescope observed the anomaly with the 47 minute cadence as the regular survey mode.
Because the anomaly was very short, $\sim$1 day, and started after the last OGLE observation, 
we could not increase the cadence nor issue the anomaly alert in a timely manner.
Nevertheless, the normal cadence is enough to reveal the perturbation caused by planet.
The OGLE data with a cadence of once per night until the anomaly, are also very important for the characterization of this event. 
In particular, the OGLE data shows us that the dip had not started by HJD$' \simeq $ 6147.6, had commenced by HJD$' \simeq $ 6148.6 and 
had almost ended by HJD$' \simeq $ 6149.6. 

This event does not cross any caustic curves and, unfortunately, 
MOA could not obtain data on HJD$' \sim  $ 6148, which corresponds to the start of the anomaly owing to bad weather.
Because of these factors, we have no data on a steep gradient of magnification in this event, thus we cannot detect a significant finite source effect.
In addition, we took AO images of the target in the year following the discovery, using the Keck telescope.
We describe the details of the Keck observations and the analysis in Section \ref{sec-KECK}.

\section{Data reductions} \label{sec-reduction}
Our data-sets for the modeling below consist of 1275 OGLE $I$-band data points, 81 OGLE $V$-band data points, 6324 MOA-Red data points and 382 B$\&$C $I$-band data points.

The OGLE data were reduced by the OGLE Difference Image Analysis (DIA) photometry pipeline \citep{uda03}.
The centroid of the catalogued star near the event, which is used for PSF photometry in the standard OGLE pipeline, 
is significantly different from that of actual event on the difference image. 
So, we rerun the PSF photometry with the real centroid for the event manually to obtain more accurate photometry.

The images taken by the MOA-II telescope and the B$\&$C telescope were reduced by the MOA DIA pipeline \citep{bon01}.
In the crowded stellar field images of the Galactic bulge, the precision of photometry is very sensitive to seeing.
We found a systematic photometry bias correlated with the seeing value in the MOA-Red data.
We reduced this systematic error by modeling it with a polynomial of seeing and airmass by using the baseline;
\begin{align}
F_{\rm cor} &= a_0 + a_1 ~{\rm JD} + a_2 ~ {\rm airmass} + a_3 ~{\rm airmass}^2 + a_4 ~{\rm seeing} + a_5 ~{\rm seeing}^2 \notag\\
 &+ a_6 ~\tan {z} \cos {\phi} + a_7 ~\tan {z} \sin{\phi} + a_8 ~{\rm airmass}\tan{z} \cos{\phi} ~{\rm seeing} \notag\\
 &+ a_9 ~{\rm airmass} \tan{z}\sin{\phi} ~{\rm seeing} + a_{10} ~{\rm airmass} \tan{z}\cos{\rm \phi}~ {\rm seeing}^2 \notag\\
 &+ a_{11} ~{\rm airmass} \tan{z}\sin{\rm \phi}~{\rm seeing}^2
\end{align}
 where $z$ and $\phi$ are the elevation angle and parallactic angle of the target included to correct the differential refraction, respectively.
$F_{\rm cor}$ is the additional flux for the correction and the corrected flux $F_{\rm new}$ is $F_{\rm new} = F_{\rm cor} + F_{\rm old}$, where $F_{\rm old}$ is the original flux from the DIA pipeline.
In the resulting photometry, the $\chi^2$ goodness-of-fit value for the time series of baseline is improved by $\Delta \chi^2 \sim  0.07$ per data point.

The relative error of data points given by the photometry code are robust for a given instrument. 
However it is known that the absolutely value of uncertainty are underestimated 
in such stellar crowded fields for various reasons in general.
Thus we empirically normalize the errors in each data-set to estimate the proper uncertainties of fitted model parameters.
We used the formula presented in \citet{yee12} for normalization, 
$\sigma _i' = k \sqrt{\sigma^2_i + e_{\rm min}^2}$
where $\sigma_i$ is the original error of the $i$th data point in magnitudes, and the parameters for normalization are $k$ and $e_{\rm min}$.
$k$ and $e_{\rm min}$ are adjusted so that the cumulative $\chi^2$ distribution as a function of the number of data points sorted 
by each magnification of the preliminary best-fit model is a straight line of slope 1. 
By including $e_{\rm min}$, we can correct the error bars at high magnification, which can be affected by flat-fielding errors.
But we found unusually large $e_{\rm min}$ values for the OGLE $I$ and MOA-Red data (0.02 and 0.09 respectively) and 
the deviations from a straight line in cumulative $\chi^2$ distribution mainly arose from baseline data points, i.e., 
not from high magnification data points as expected.
Thus it is not reasonable to normalize errors with this values and we adopt $e_{\rm min} = 0$ for these two data-sets.
This may indicate that there is some low-level systematics in the light curve.
We apply $e_{\rm min} = 0,~ k = 1.364$ to OGLE $I$, $e_{\rm min} = 0,~ k = 1.576$ to OGLE $V$, $e_{\rm min} = 0,~ k = 0.907$ to MOA-Red and 
$e_{\rm min} = 0.0061, k = 1.021$ to B$\&$C $I$. 
The normalization factors applied for the OGLE $I$ and $V$ data are consistent with those given by \citet{sko16a}.
Note that the final best fit model parameters are consistent with the preliminary model parameters before the error normalization.
Thus this procedure of the error normalization does not affect out main result.
The parameters of these data sets are also shown in Table \ref{tab-data}.

\section{Modeling} \label{sec-model}
Here we present and compare the results of our light curve modeling assuming a standard binary lens and also
adding the effects of parallax.
We fit the light curves using a Markov Chain Monte Carlo (MCMC) approach \citep{ver03}, with magnification calculations from 
image centered ray-shooting method \citep{ben96,ben10}.

\subsection{Standard binary lens}
In the case of a point lens,  the magnification map on the source plane is circular symmetric around the lens.
In the point source point lens (PSPL) case, we can characterize the microlensing light curve with the time of the closest approach of the source to the center of mass of the lens, $t_0$,
the minimum impact parameter $u_0$ at $t_0$, and the Einstein radius crossing time (or timescale) $t_{\rm E}=\theta_{\rm E}/\mu_{\rm rel}$,
where $u_0$ is in units of $\theta_{\rm E}$ and $\mu_{\rm rel}$ is the lens-source relative proper motion.
When the lens has a companion, its the gravity distort the magnification map and create the closed curves called as caustics where the magnification is infinite.
In this case, three parameters are added to the fitting parameters above; the mass ratio of two lenses $q$ and their angular separation normalized 
by $\theta_{\rm E}$, $s$, which determine the shape and location of the caustics, and the source trajectory with respect to the binary lens axis, $\alpha$,
which determines the direction of a one-dimensional slice of the distorted magnification map.
When a source star crosses a region with a steep gradient near the caustics in the magnification map, we can observe the finite source effect.
Because source stars of most binary lens events cross such regions, 
we include the source size $\rho \equiv \theta_*/\theta_{\rm E}$ as a fitting parameter for a binary lens model.
With the magnification variation against time, $A(t,\bm{x})$, 
which is defined in terms of the above parameters $\bm{x}=(t_0,u_0,t_{\rm E},q,s,\alpha,\rho)$, we can linearly fit
\begin{equation}
F(t) = f_S A(t,\bm{x}) + f_b \label{eq-F}
\end{equation}
to a data set and obtain the instrumental source flux $f_S$ and the instrumental blending flux $f_b$ for every telescope and pass-band.

We adopt a linear limb-darkening law with one parameter, $u_{\lambda}$.
According to \citet{gon09}, we estimate the effective temperature, $T_{\rm eff}\sim 5500$ K from the source color which is discussed in 
Section \ref{sec-color} and assumed the solar metallicity. With $T_{\rm eff} = 5500$ K and assuming surface gravity $\log~ g = 4.0~ {\rm cm~ s^{-2}}$
and microturbulence parameter $\xi = 1.0~ {\rm km ~s^{-1}}$, the limb-darkening coefficients selected from \citet{cla00} are $u_I=0.5470$ for 
OGLE $I$ and B$\&$C $I$, $u_V = 0.7086$ for OGLE $V$, and $u_{\rm MOA-Red} = 0.5895$ for MOA-Red which is the average of standard $I$ and $R$ filters.
Therefore we used the $u_I$ for OGLE $I$ and B$\&$C $I$, the $u_V$ for OGLE $V$ and the mean of the $u_I$ and $u_R$, 0.5895 for MOA-Red, 
the filter which has the range of both the standard $I$ and $R$ filters.
Although the best estimated value of $T_{\rm eff}$ and the limb-darkening coefficients depend on the source magnitude in each model,
we keep using the fixed values.
However, this does not affect our final result because the finite source effect is very weak in the light curve, as mentioned below.
The limb-darkening coefficients we adopt are also shown in Table \ref{tab-data}.

We show the parameters of the best-fit models of our standard binary lens modeling in Table \ref{tab-models}, 
where the uncertainties shown are from the 16th/84th percentile values of the stationary distributions given by MCMC.
We find a degeneracy between the close model of $s < 1$ and the wide model of $s > 1$ with $\Delta \chi^2 \simeq 0.7$.
There is a well-known degeneracy in high magnification microlensing events between lens systems with similar mass ratios, 
but separations $s$ and $1/s$.
In microlensing events suffering this degeneracy, the source star passes close to the central caustic which has, 
in each of the degenerate solutions, a similar shape.
However, in this event, the close/wide degeneracy has a different nature, in terms of the caustic geometry.
A single resonant caustic is seen in the wide model with $s = 1.007$ while the caustic curves are separated into central caustic and planetary caustics in the close model 
with $s = 0.890$.
As seen in Figure \ref{fig-caus}, it is understood that the gradients of magnification on the source trajectories are similar in both models although the caustic shapes are different.
The mass ratios are $q \simeq 2 \times 10^{-4}$ in both models indicating the companion has a planetary mass.
We find that the finite source effect is weak and the $\rho$ value is consistent with $\rho = 0$ at the 1 $\sigma$ level. 
Because a larger $\rho$ value reduces the dip depth in the light curve and does not explain the data, we can place an upper limit on $\rho$.

\subsection{Parallax}
In long timescale microlensing events, such as this one, the effect of Earth's orbital motion around the Sun may be detectable \citep{gou92,alc95}.
This effect is expressed by the microlens parallax vector $\bm{\pi_{\rm E}} = (\pi_{\rm E,N}, \pi_{\rm E,E}) = \pi_{\rm E}~ \bm{\mu_{\rm rel}}/\mu_{\rm rel}$ \citep{gou00a}.
Here, $\pi_{\rm E,N}$ and $\pi_{\rm E,E}$ are the north and east components of $\bm{\pi_{\rm E}}$, respectively, whose direction is same as lens-source relative proper motion.
The magnitude $\pi_{\rm E} \equiv 1{\rm AU} / \tilde{r}_{\rm E}$, is defined by 
1 AU relative to the Einstein radius projected onto the observer plane $\tilde{r}_{\rm E} = R_{\rm E} D_S/(D_S - D_L)$.

\begin{figure}
\centering
\epsscale{0.6}
\rotatebox{-90}{
\plotone{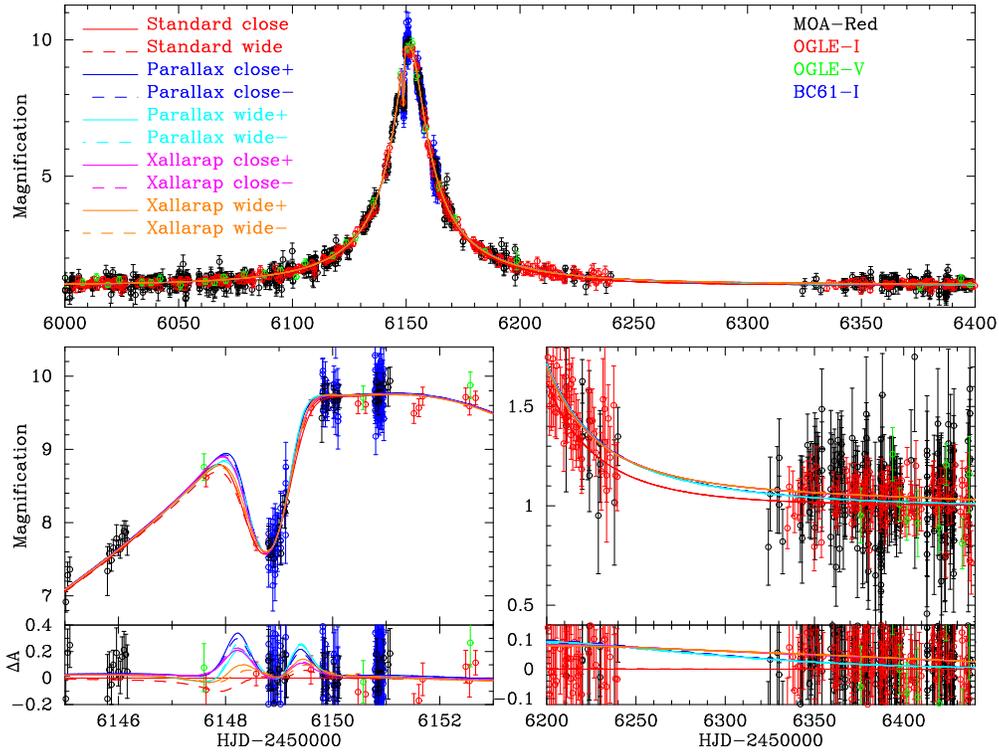}
}
\caption{The light curve of OGLE-2012-BLG-0950 with the best-fitting models indicated in the top left.
The top panel shows the whole event, the bottom left and the bottom right panels highlight the planetary anomaly and 
the light curve from the end of 2012 to the start of 2013, respectively.
The residuals from the Standard close model are shown in the bottom insets of the bottom panels.}
\label{fig-models}
\end{figure}

\begin{figure}
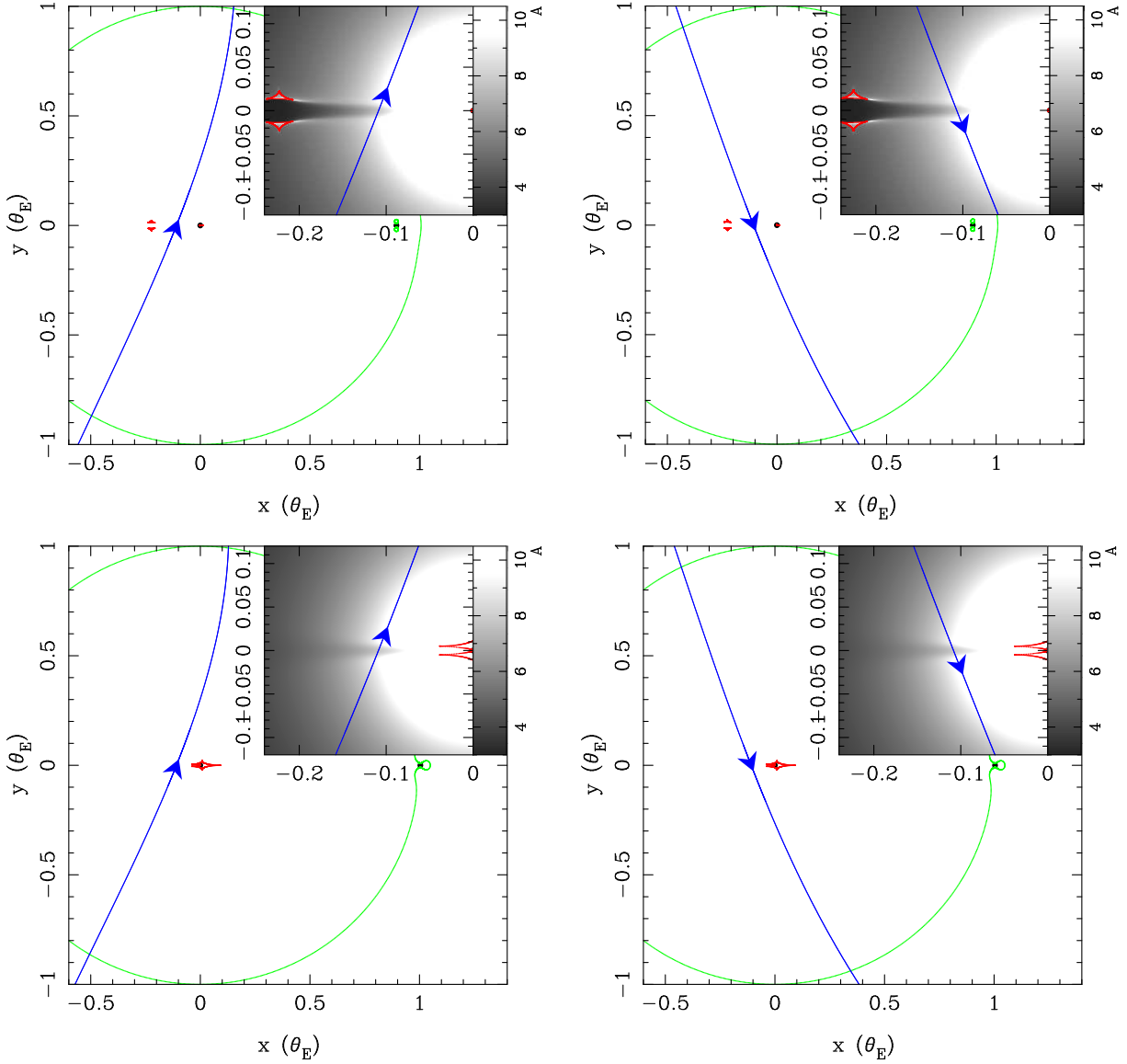

\begin{minipage}{0.5\hsize}
\begin{center}
\rotatebox{-90}{
\includegraphics[width=75mm]{parac+cau.eps}
}
\end{center}
\end{minipage}
\begin{minipage}{0.49\hsize}
\begin{center}
\rotatebox{-90}{
\includegraphics[width=75mm]{parac-cau.eps}
}
\end{center}
\end{minipage}
\begin{minipage}{0.5\hsize}
\begin{center}
\rotatebox{-90}{
\includegraphics[width=75mm]{paraw+cau.eps}
}
\end{center}
\end{minipage}
\begin{minipage}{0.49\hsize}
\begin{center}
\rotatebox{-90}{
\includegraphics[width=75mm]{paraw-cau.eps}
}
\end{center}
\end{minipage}
\caption{Caustics for the parallax models. The blue arrowed lines indicate the source trajectories.
The top left, top right, bottom left and bottom right shows the parallax close$+$, close$-$, wide$+$ and wide$-$ models, respectively.
A magnification map around the anomaly part is shown in the inset of each panel.}
\label{fig-caus}
\end{figure}

We show our best-fit parallax models in Figure \ref{fig-models}, \ref{fig-caus} and Table \ref{tab-models}.
We found each close and wide solutions has an additional degeneracy between $u_0 > 0 (+)$ and $u_0 < 0 (-)$.
These four degenerate solutions have $\Delta \chi^2 \lesssim 4$.
The parameters of the degenerate models are consistent with each other to within $1\sigma$ error except for $s$, $\alpha$ and $u_0$.
The $\chi^2$ difference between the standard models and the parallax models is significantly large, $\Delta \chi^2 > 110$ for 2 dof difference.

It is known that low-level systematics in the baseline sometime mimic a high order signal.
We therefore check whether the $\Delta \chi^2$ contributions come from where we theoretically expect.
The top inset in Figure \ref{fig-para-stand} shows the cumulative distribution for $\Delta \chi^2$ between the Standard close model and 
the parallax close$+$ model as a function of time.
Positive $\Delta \chi^2$ values indicate that the parallax model is favored over the standard model.
We find $\Delta \chi^2 \sim 90$ comes from the data during the main peak of event in 2012 in both MOA and OGLE as expected,
and $\Delta \chi^2 \sim 25$ comes from the data in 2013, the next year.
The bottom right panel of Figure \ref{fig-models}, which is a zoom of the 2013 data, shows slight differences among the models, i.e., 
the parallax models have the magnification of $\sim 1.05$ in the start of 2013 while the standard models have $\sim 1.00$.
The bottom panel in Figure \ref{fig-para-stand} shows binned residuals in bins 25 days wide.
This shows the clear long term deviation from the standard model. 
The binned data in the first half of 2013 are mostly above the standard model in both MOA and OGLE whereas those in the other years are not,
which can be well explained by the the parallax model.
Because these long term distortion are consistent in both MOA and OGLE,
we conclude that this long term signature are real and they are better explained by the parallax models compared to the standard models.
Note that adding lens orbital motion does not improve our models. We also modeled the orbital motion of the source star due to the companion,
an effect called xallarap, and conclude that the parallax scenario is preferred over the xallarap scenario.
See Tables \ref{tab-models} and \ref{tab-property} for the modeling results, and the additional details in the Appendix.
We consider only the parallax model in the following sections.

\section{The Angular Einstein Radius} \label{sec-color}
We can place an upper limit on $\rho$ for the parallax models and a lower limit as well for the xallarap models.
It is possible to derive a constraint on $\theta_{\rm E} = \theta_*/\rho$ by obtaining the angular source radius $\theta_*$.
$\theta_*$ can be estimated from the source color, $(V - I)_S$, and the magnitude, $I_S$, empirically.
We used the empirical relation by using a result of \citet{boy14} analysis,
\begin{equation}
\log ~ [2 \theta_{*}/(1 {\rm mas})] = 0.5014 + 0.4197 (V-I) - 0.2 I. \label{eq-boya} 
\end{equation}
This relation comes from a private communication with them, which is restricted to FGK stars with 3900 K $< T_{\rm eff} <$ 7000 K and the accuracy of relation is better than 2$\%$ \citep{fuk15}.
We measured the source color and brightnesses $(V-I, I)_S = (1.346, 19.29) \pm (0.001, 0.03)$ with the OGLE-IV instrumental magnitude from the light curve fitting.
Note that because the source color and magnitudes are nearly identical for all models (see Table \ref{tab-models}), 
we adopt values of parallax close$+$ model in the following analysis.
We correct their extinction following the standard procedure by \citet{yoo04} using the red clump giants (RCG) as a standard candle.
Figure \ref{fig-cmd} shows a color-magnitude diagram (CMD) within the $2' \times 2'$ region around the source star with the OGLE-IV instrumental magnitude.
The position of source and the measured RCG centroid $(V-I, I)_{RC} = (1.644, 15.27) \pm (0.011, 0.04)$ are shown as blue dot and red cross, respectively.
Comparing the measured RCG centroid and the expected intrinsic position in this field $(V-I, I)_{RC,0} = (1.06, 14.38) \pm (0.07, 0.04)$ 
by \citet{ben13} and \citet{nat13}, we obtain the intrinsic source color and magnitude as $(V-I, I)_{S,0} = (0.76, 18.40) \pm (0.07, 0.07)$ 
with the assumption that the source extinction are same as that the RCG.
Note that the original reddening and extinction values in the standard magnitude in this field can be measured by 
the catalog of OGLE-III photometry map \citep{szy11} as $E(V-I) = 0.68 \pm 0.07$ and $A_{I} = 0.86 \pm 0.06$. We use these original values to obtain $A_H$ in Section \ref{sec-KECK}.
Applying the intrinsic source color and magnitude to Equation (\ref{eq-boya}), we obtain the angular source radius as $\theta_* = 0.69 \pm 0.05 \ \mu {\rm as}$.
From $\theta_*$, $\rho$ and $t_{\rm E}$, we can calculate the angular Einstein radius $\theta_{\rm E}$ and the lens-source relative proper motion $\mu_{rel}$, 
\[ \theta_{\rm E} = \theta_*/\rho > 0.22 \ {\rm mas}, \ \  \mu_{rel} = \theta_{\rm E}/t_{\rm E} > 1.2 \ {\rm mas / yr} \]
for the parallax close$+$ model.
Table \ref{tab-property} shows the derived parameters for all degenerate models using each models' values.

\section{Excess Brightnesses from Keck AO Images} \label{sec-KECK}
We have a mass-distance relation via the microlens parallax $\pi_{\rm E}$.
If we can measure the lens flux which gives us an additional mass-distance relation, we could measure the mass and distance of the lens uniquely.
We conducted high angular resolution observations using adaptive optics in order to measure the lens flux excluding 
as much flux from unrelated stars as possible.

\subsection{Observations and the photometry}
We observed OGLE-2012-BLG-0950 with the NIRC2 instrument on Keck II on July 18, 2013.
We used the Wide camera with a pixel scale of 0.04 arcsec. We took 15 dithered $H$ frames with an exposure
time of 30 seconds. We performed dark and flatfield corrections in the standard way.
Furthermore, OGLE-2012-BLG-0950 was observed as part of the VVV survey \citep{min10} using the VISTA 4m telescope at ESO.
We extracted 3 arcmin VVV $JHK$ images centered on the target.  We used the suite of tools developed as part of astr$O$matic \citep{ber02}.
We analysed the PSF of the images using PSFEx, then we measured the fluxes with SExtractor \citep{ber96} using these PSF models.
We cross identified 2MASS stars with VVV sources, and derived an absolute calibration of the VVV $JHK$ images.
We used the VVV images which we reprocessed as a reference to perform an astrometric calibration of one Keck frame.
We then extracted the sources from this individual frame, and used them as a reference to realign all the Keck frames.
We stacked the Keck frames with the SWARP tool \citep{ber02}. We then performed aperture photometry on the Keck frame (for more details,
see Batista et al. 2014). We cross identified common sources between Keck and VVV, and finally derived the calibration 
constant for Keck $H$ band photometry.

Fig. \ref{fig-KECK} shows the field as observed by VVV and by Keck. First, we notice that the source in VVV is resolved in 2 objects with Keck.
Using the astrometry on the amplified source, we are able to identify the source$+$lens of the microlensing as being the star that is
marked on the frame. Its coordinates are $(R.A., Dec.)_{J2000}$ = (18:08:04.620, -29:43:53.43). It has a Keck $H$ band magnitude of 
\begin{equation}
H_{\rm target} = 16.89 \pm 0.02 \label{eq-HKECK}
\end{equation}
in the 2MASS magnitude system. 
The blend to the south is at coordinates $(R.A., Dec.)_{J2000}$ = (18:08:04.612, -29:43:53.88) and is slightly fainter at $H=16.99 \pm 0.03$.

\subsection{The excess flux} \label{sec-excess}
Considering a full width at half maximum of the target FWHM$_{\rm target} =$ 90 mas in Keck image and the lens-source relative proper motion,
$H_{\rm target}$ includes the lens flux plus the source flux.
Here we derive the excess brightness by subtracting the source system brightness, which can be used as the lens flux or its upper limit.
Hereafter we represent our derived values for only the parallax close$+$ model as the parallax model unless otherwise stated.

Unfortunately, we don't have a light curve in the $H$-band, so we derived the source $H$ magnitude as $H_{S,0} = 17.55 \pm 0.12$ by converting the magnitude from 
$(V-I, I)_{S,0}$ with the color-color relation by \citet{bes88}.
Next we applied the $E(V-I)$ and $A_I$ values derived in Section \ref{sec-color} and estimated the extinction in $H$-band as $A_H = 0.25 \pm 0.02$ \citep{car89}.
The magnification at the time of Keck observation at HJD$'$ = 6491.88 is $A = 1.005$ for the parallax model.
Thus the source apparent $H$ magnitude at the time is $H_{S, {\rm KECK}} = 17.78 \pm 0.12$. This value is converted to the 2MASS magnitude system from the Bessel $\&$ Brett system using Equations (A1) - (A4) in \citet{car01}.
Subtracting this from $H_{\rm target}$ of Equation (\ref{eq-HKECK}), we derive the excess brightness of
\begin{equation}
H_{\rm excess} = 17.52 \pm 0.10. \label{eq-expara}
\end{equation}

\subsection{Probability of the contamination fraction $f$} \label{sec-probcont}
We next consider the probability that part of the $H_{\rm excess}$ come from stars other than the lens.
To estimate the probability of the contamination fraction $f$, we consider the following three possible sources of contamination: unrelated ambient stars, 
a companion to the source star and a companion to the lens star \citep{bat14, fuk15}.

\subsubsection{Unrelated ambient stars}
First, we estimate the probability of contamination owing to unrelated ambient stars.
Counting the number of stars with $ H > H_{\rm excess}$ on the Keck image, we estimate the number density of stars with brightnesses corresponding to $0 < f < 1$ as 0.68 arcsec$^{-2}$.
Similarly, the number density of stars corresponding to $0.5<f<1$ and $0.9<f<1$ are estimated as 0.11 arcsec$^{-2}$, 0.01 arcsec$^{-2}$, respectively.

We can resolve an ambient star only if it is separated from the source by FWHM = 90 mas or more.
Thus the probability of contamination owing to unrelated ambient stars within 90 mas around the source, $P_{amb}$, are
$P_{amb}(0<f<1) =  1.74 \%$, $P_{amb}(0.5<f<1) =  0.29 \%$ and $P_{amb}(0.9<f<1) =  0.03\%$.

\subsubsection{Companion to the source star}
Second, we estimate the probability of contamination owing to a companion to the source.
Because we can detect a companion on the Keck images if the companion is located far enough from the source and the light curve will be affected if the companion is 
located close enough to the source, an undetectable companion should be located between the two detection limits.
We put the distant limit as 90 mas from the FWHM of the target.
For the close limit, we consider $a_{SC,{\rm low}}$ as defined in the following.

As we can find in the bottom inset in Figure \ref{fig-para-stand}, the OGLE $I$ data in the light curve is sensitive enough to
a deviation with an amplitude of $\Delta A \simeq 0.1$ and a duration of hundreds of days.
That means we can detect a binary source signal when a companion to the source is magnified with an amplitude of $\gtrsim 0.1 f_S$ and a duration of hundreds of days.
Such small signals cannot be detected if they are longer than $10t_{\rm E} \sim 670 {\rm ~days}$. We assume that we can detect the signal caused by a magnified companion 
to the source star when the time variation of the companion flux is larger than $0.1 f_S$ within $10 t_{\rm E}$, that is, it requires
\begin{align}
A(u_{C,0}) f_C - A\left(\sqrt{5^2+u_{C,0}^2}\right) f_C \leq 0.1 f_S
\end{align}
to be "undetected", where 5 comes from half of "10" $t_{\rm E}$ and $u_{C,0}$ and $f_C$ are the impact parameter and the flux in the $I$-band of the companion star, respectively.  $A(u_C)$ is the magnification when the companion is located at $u_C$ and we use the magnification owing to a single lens $A(u) = (u^2+2)/u\sqrt{u^2+4}$ as $A(u_C)$.
Because $A\left(\sqrt{5^2+u_{C,0}^2}\right) < A(5)$, we can express the condition more conservatively,
\begin{align}
A(u_{C,0}) f_C - A(5) f_C \leq 0.1 f_S\notag \\
\Leftrightarrow u_{C,0} \geq u_C \left( 1.00275+0.1\frac{f_S}{f_C} \right) \equiv u_{C,0,{\rm low}}. \label{xalla-criU}
\end{align}

The $u_{C,0,{\rm low}}$ is a lower limit of the impact parameter of the companion to be undetected and it gives us a lower limit of the separation between 
the source and the companion as $s_{SC} > s_{SC,{\rm low}}$.
When the time when the companion is located at $u_C = u_{C,0}$ is defined as $t_{C,0}$, it is possible to give the most conservative lower limit as 
$s_{SC} > s_{SC,{\rm low}} \equiv |u_{C,0,{\rm low}} - |u_0||$ when $t_{C,0} = t_0$ and the closest companion crosses the same side as the source relative to the lens.
Thus we derive the lower limit of the semi-major axis of the source system as $a_{SC,{\rm low}} \equiv D_{S} \theta_{\rm E} s_{SC,{\rm low}} $.

To calculate $u_{C,0,{\rm low}}$ in $a_{SC,{\rm low}}$, we obtain $M_{I,C}$ by converting $M_{H,C} = H_{\rm excess} - 2.5\log{f} -5\log{(D_S/10{\rm pc})}$ to $M_{I,C}$ 
using PARSEC isochrones version 1.2S \citep{bre12,che14,che15,tan14} with $D_S = 8$kpc for the calculation of $f_S/f_C = 10^{-0.4(M_{I,S} - M_{I,C})}$ in Equation (\ref{xalla-criU}) 
where $M_{I,S}$ and $M_{I,C}$ are absolute $I$ magnitudes of the source and the companion.

Because $a_{SC,{\rm low}}$ and $\theta_{\rm E}$ vary with $f$, we calculate the probability of contamination owing to a companion to the source using following formula \citep[cf.]{fuk15},
\begin{equation}
P (f_1<f<f_2) = F_{\rm binary} \times \int_{f_1}^{f_2} F_{a_c}(f) \times F_{q_c}(f) df \label{eq-probc}
\end{equation}
where $F_{\rm binary}$ is the multiplicity of FGK-dwarfs, $F_{a_c}(f)$ is the fraction of binaries with a separation of $a_{SC,{\rm low}}(f) <a_c< 90{\rm mas} \times D_S$
and $F_{q_c}(f) df$ is the fraction of binaries with a mass ratio between $q_c(f)$ and $q_c(f+df)$.
We derive $M_C(f)$ from $M_{H,C}(f)$ using the PARSEC isochrones version 1.2S for the calculation of $q_c(f)$.
We use the distribution of parameters for FGK binaries by \citet{rag10}: $F_{\rm binary} = 0.46$ as the multiplicity, a log normal distribution with 
the mean of $\log P_c({\rm days})=5.03$ and the standard deviation of $\sigma_{\log P_c}=2.28$ as the period distribution and Figure 16 in the paper 
as the mass ratio distribution.
We apply the period distribution by converting $a_c$ to a period using Kepler's 3rd law.
Then we obtain $P_{SC}(0<f<1) = 22.0 \%$, $P_{SC}(0.5<f<1) = 3.2 \%$, $P_{SC}(0.9<f<1) = 0.7 \%$ for the parallax close$+$ model.

\subsubsection{Companion to the lens star}
Next, we estimate the probability for contamination owing to a companion to the lens star.
In this case, we again use Equation (\ref{eq-probc}) to calculate the probability because the lens mass, the distance to the lens and the companion mass values are 
dependent on the $f$ value in addition to $\theta_{\rm E}$.
However, the definition of $F_{a_c}(f)$ is different here.
To place close limits on $a_c$, we consider the shear given by a hypothetical lens companion \citep{bat14}.
Assuming that we can detect the shear effect when the width of the central caustic created by the companion, $w_c \simeq 4q_c/s_c^2 $, is larger than 
the width of that by the planet, $w$, we can place a detection limit as $w_c < w$, where $s_c$ is the projected separation normalized by $\theta_{\rm E}$ between 
the host and a hypothetical companion of the lens. This inequality gives us a close limit on $a_c$. 

Adopting 90 mas as the distant limit again, $F_{a_c}(f)$ here is defined as 
the fraction of binaries with a projected separation of $2\sqrt{q_c(f)/w} \theta_{\rm E}(f) D_L(f) < a_c < 90 {\rm mas} \times D_L(f)$ where 
$w = 4q/(s-s^{-1})^2 \simeq 0.016 \sim 0.02$ for the close models. 
We apply $w=0.02$ to the wide model as well as the close model considering that the shear effect by a hypothetical companion 
is almost equal to that for the close model because the magnification map on the source trajectory for the wide models is almost same as the close models.
Note that uncertainties arising from any assumptions here do not affect the results largely because they are too small compared to the range of $a_c$ values we are considering.

We calculate $q_c(f)$ for each $f$ value using isochrone models as well as the source companion case.
Because a primary of the lens can be either an M dwarf or FGK dwarf depending on the $f$ value, we use the distributions for FGK-dwarf binaries by \citet{rag10} 
for $M_{\rm prim}(f) > 0.7$ and that for KM-dwarf binaries by \citet{war15} for $M_{\rm prim}(f) \leq  0.7$. 
\citet{war15} gives $F_{\rm binary} = 0.347$ as the multiplicity, a log normal distribution with the mean of $\log a_c ({\rm AU}) = 0.77$ and the standard deviation of
$\sigma_{\log a_c} = 1.34$ as the distribution of the projected separation and a mass ratio distribution that is flat for $0.2<q_c<1$ and 0 for $q_c<0.2$.
Integrating Equation (\ref{eq-probc}) with these distributions and the detection limits, we derive $P_{LC}(0<f<1) = 13.7 \%$, $P_{LC}(0.5<f<1) = 7.2 \%$ and 
$P_{LC}(0.9<f<1) = 3.9 \%$ for the parallax close$+$ model.

\subsubsection{Total probability of contamination}
Finally, by summing the probabilities for the three sources mentioned above, we can calculate the following total probabilities of contamination:
$P(0<f<1) =  37.4 \%$, $P(0.5<f<1) = 10.7\%$, $P(0.9<f<1) = 4.6\%$ assuming the parallax close$+$ model. 
Therefore, the probabilities of the contamination fraction not exceeding $f$ shown in Table \ref{tab-property} can be calculated as $P(f = 0) = 1 - P(0<f<1) = 62.6 \%$,
$P(f\leq 0.5) = 1 - P(0.5<f<1) = 89.3 \%$ and $P(f\leq 0.9) = 1 - P(0.9<f<1) = 95.4 \%$ for the parallax close$+$ model.
The probabilities for all parallax models are shown in Table \ref{tab-property}.

\section{Lens Properties} \label{sec-lens}
In this section, we constrain the lens properties from microlens parallax and the lens flux.
All results are summarized in Table \ref{tab-property}.

We have two mass-distance relations for the parallax model.
One is the lens absolute magnitude,
\begin{equation}
M_{H,L} = H_L - A_{H,L} - 5 \log{\frac{D_L}{10 {\rm pc}}} \hspace{0.5cm} (H_L= H_{\rm excess} - 2.5 \log{(1-f)}) \label{eq-mhdl1}
\end{equation}
where $f$ is the fraction of contamination flux to excess flux and $A_{H,L}$ is an extinction for the lens located at $D_L$ and
we adopted $A_{H,L} = A_H D_L/D_S$ following to \citet{fuk15}.
This is converted to a relationship between the host mass $M_{\rm host}$ and the distance $D_L$ by using isochrone models from 
the PARSEC isochrones, version 1.2S \citep{bre12, che14, tan14, che15}.
The other mass-distance relation is from the microlens parallax $\pi_{\rm E}$:
\begin{align}
M_{\rm host} &= \frac{1}{1+q} \frac{\pi_{\rm rel}}{\kappa \pi_{\rm E}^2} \label{eq-mhdl2}
\end{align}
where $\kappa = 8.144 ~{\rm mas} ~M_\odot ^{-1}$ and $\pi_{\rm rel} = {\rm AU}~ (1/D_L - 1/D_S)$.

Figure \ref{fig-MhDL-para} shows these two relations for the parallax close$+$ model.
Black lines are the mass-distance relation come from $\pi_{\rm E}$.
The red, blue and green lines indicate the relation from $M_{H,L}$ for the case of $f$=0, 0.5 and 0.9, respectively.
Dashed lines indicate 1 $\sigma$ errors and they include the uncertainty of the distance to the source of $D_S=8.0 \pm 1.6$ kpc, the lens age of $<$ 13 Gyr and 
the lens metallicity of [Fe/H] = $-0.05 \pm 0.20$ in addition to the uncertainty of our measurements. 
We adopt the metallicity distribution of nearby M- or late K-dwarf stars \citep[e.g.]{gai14} for the metallicity.
Note that the dependency on age is much weaker than that on metallicity in the region of the parallax solution.

The region overlapping these two relations corresponds to the allowed solution.
For $f=0$, the host mass is $M_{\rm host} = 0.57^{+0.06}_{-0.10} M_{\odot}$ and 
the distance is $D_L = 2.6^{+0.5}_{-0.7}$ kpc, and the planet mass is $M_{\rm p} = 35^{+10}_{-6} M_{\oplus}$, its projected separation is 
$r_{\perp} = 2.6^{+0.3}_{-0.5}$ AU and the three-dimensional star-planet separation is statistically estimated as $a = 3.1^{+1.5}_{-0.7}$ AU with 
a circular orbit assumption \citep{gouloe92}.
From this solution, we can calculate the angular Einstein radius and the relative lens-source proper motion as $\theta_{\rm E}= 1.09^{+0.16}_{-0.10}$ mas and 
$\mu_{\rm rel} = 5.8^{+0.8}_{-0.6}$ mas/yr, respectively.
The solution of $M_{\rm host} = 0.50^{+0.05}_{-0.09} M_{\odot}$, $D_L = 2.9^{+0.6}_{-0.8}$ kpc for $f=0.5$ 
is consistent with that for $f=0$ within 1$\sigma$.
In addition, the solution for $f=0.9$ is $M_{\rm host} = 0.33^{+0.05}_{-0.08} M_{\odot}$, $D_L = 3.6^{+0.6}_{-0.8}$ kpc.
Our estimate for these contamination probabilities is discussed in  Section \ref{sec-probcont} and summarized also in Table \ref{tab-property}.

In any case, the host star is an M/K dwarf and the planet is a  Neptune/sub-Saturn mass planet.
All solutions of the other degenerate parallax models are similar to these results as shown in Table \ref{tab-property}.
We present the mean value of the 8 parallax solutions with $f = 0$ and $f = 0.5$ without any weight as "Mean" in Table \ref{tab-property}.
Here the contributions of solutions with $f>0.5$ are negligible.


\section{Discussion and Conclusion} \label{sec-disc}
We analyzed the microlensing event OGLE-2012-BLG-0950. 
A negative perturbation in the microlensing light curve consistent with a low-mass planet was detected \citep{abe13}.
All the models we analyzed have a planetary mass ratio, $q \simeq 2 \times 10^{-4}$.
We could not detect a significant finite source effect because the source did not cross any caustic,   but we did detect a parallax signal.
The parallax solutions indicate a Neptune/sub-Saturn mass planet with mass of $M_{\rm p} = 35^{+17}_{-9} M_{\oplus}$  around an M/K-dwarf host with mass of $M_{\rm host} = 0.56^{+0.12}_{-0.16} M_{\odot}$.
We measured the lens mass by combining microlens parallax and the lens flux obtained by Keck AO observations.
This is the first case in which the lens mass was measured using only microlens parallax and the lens flux.

The planet orbits outside of the snow line of the host star and has 
a mass between that of Neptune and Saturn, $M_{\rm p} = 35_{-9}^{+17} M_{\oplus}$.
Planets with this mass range (intermediate mass, hereafter) are predicted to be rare inside the snow line, 
but to be common like Neptune- or Saturn- mass planets outside the snow line according to the core accretion theory \citep{ida04, ida13}.
A paucity of intermediate mass planets orbiting close to their metal-poor host stars is confirmed \citep{bea13}.
On the other hand,  the predicted relative abundance outside the snow line has not been confirmed yet.
Figure \ref{fig-exo} shows the distribution of the exoplanets\footnote{http://exoplanet.eu} discovered so far.
The solution of our parallax model is indicated as the purple filled circle located just around the valley of the bimodal mass distribution histogram on
the left side of the figure.
Note that this distribution is not corrected for detection efficiency.
Only a few intermediate mass planets orbiting outside the snow line have been discovered by the RV and microlensing methods.
The parallax model of this work could be the second such intermediate mass exoplanet with mass measurement, following OGLE-2012-BLG-0026Lb \citep{han13,bea16}.
The mass of other intermediate mass planets are estimated by Bayesian analysis \citep{miy11,pol14,sko16b}.
However the Bayesian estimates depend on the choice of prior \citep{ben14,sko15}.

In a future space-based microlensing survey by {\it WFIRST} \citep{spe15} or {\it Euclid} \citep{pen13},
and in the survey of Campaign 9 of the {\it K2} Mission \citep{hen15} conducted from April 2016 to July 2016,
or the {\it Spitzer} microlensing campaign from 2014 \citep{yeeet15},
it is important and easier to determine the lens mass for each event by combining microlens parallax and lens flux 
as pointed out by \citet{yee15} for the following reasons.
First, space- and ground-based simultaneous observations are expected to obtain microlens parallax for a significant fraction of events regardless of number of the lens bodies,
in contrast to the finite source effect which can be obtained only by observing the peak of high-mag event or caustic crossing.
Second, for low-mass and nearby lenses, the mass-distance relations derived from flux and from $\theta_{\rm E}$ are partially degenerate 
(see Figure 2 in Yee 2015 or Figure 7 in Fukui et al. 2015)
although we can obtain $\theta_{\rm E}$ by the measurement of astrometric microlensing effects with the precision of {\it WFIRST} \citep{gou14}.
Third, the cases without detection of $\theta_{\rm E}$ like this event are expected to increase even for planetary events because a higher precision and higher cadence survey can 
detect more subtle planetary signals including cases without crossing caustics \citep{zhu14}.
Finally, it is possible to measure the lens fluxes even after the events by follow-up observations with high angular resolution, and ultimately, 
{\it WFIRST} and {\it Euclid} can routinely measure the lens fluxes as part of the survey observations.
Our analysis for the parallax model is the first demonstration of the mass measurement from only microlens parallax and the lens flux, 
and thus it has particular significance for the developing era of space-based microlensing.

We acknowledge the following support: Work by N.K. is supported by JSPS KAKENHI Grant Number JP15J01676. 
The MOA project is supported by the grant JSPS25103508 and 23340064.
N.J.R is a Royal Society of New Zealand Rutherford Discovery Fellow.
OGLE Team thanks Profs.\ M.~Kubiak and G.~Pietrzy{\'n}ski, former members of the OGLE team, 
for their contribution to the collection of the OGLE photometric data over the past years.
The OGLE project has received funding from the National Science Centre,
Poland, grant MAESTRO 2014/14/A/ST9/00121 to AU.
V.B. was supported by the CNES and the DIM ACAV, R\'egion \^lle-de-France.
V.B., J.P.B., and J.B.M. acknowledge the support of PERSU Sorbonne Universit\'e, the Programme National
de Plan\'etologie and the labex ESEP.

\appendix

\section{Xallarap analysis} \label{sec-xallaap}
If the source star is in a binary system, the orbital motion of the source star can also measurably affect 
the trajectory of the source during a microlensing event.
This effect, called xallarap, requires additional parameters which define orbital elements of the source system whereas 
we know the Earth's orbital elements for the parallax effect.
This model requires 7 additional fitting parameters to the standard binary model, the direction toward the Earth relative to the source orbital plane, 
$R.A._{\xi}$ and $Dec._{\xi}$, the orbital period $P_{\xi}$, the orbital eccentricity $\epsilon$ and the perihelion time $t_{\rm peri}$ 
in addition to $\bm{\xi_{\rm E}} = (\xi_{\rm E,N}, \xi_{\rm E,E})$ which is analogous to $\bm{\pi_{\rm E}}$ for microlens parallax.
We omitted $\epsilon$ and $t_{\rm peri}$ as fitting parameters by assuming a circular orbit.
Here we describe the results of our xallarap analysis and comparison with the parallax model, and discuss the possibility.

\subsection{Constraint by the companion mass upper limit} \label{sec-chi2orb}
Kepler's 3rd law gives us a relation of the source orbit;
\begin{align}
\xi_{\rm E} = \frac{{\rm AU}}{D_S \theta_* / \rho} \left( \frac{M_C}{M_{\odot}} \right) \left( \frac{M_{\odot}}{M_S+M_C} \frac{P_{\xi}}{1{\rm yr}} \right)^{2/3} \label{eq-xiE}
\end{align}
where $M_S$ and $M_C$ are the masses of the source and its companion, respectively.
The $\xi$, $\rho$ and $P_{\xi}$ are fitting parameters, $\theta_*$ is measured, and $D_S$ and $M_S$ are reasonably constrained 
by the Galactic model combined with the source color and magnitude values. Then, for a given MCMC chain, $M_C$ can be calculated from this relation.

In xallarap fitting with no constraints, we found that the light curve prefers a solution with 
an unrealistically massive source companion with a mass of $M_C \sim 400~ M_{\odot}$.
Thus we conducted the fitting with the following constraint.
We can place an upper limit of the companion mass $M_C < M_{C,{\rm max}}$ from Equation (\ref{eq-expara}) of the excess flux 
with an assumption that the companion is not a stellar remnant.
That corresponds to placing an upper limit of $\xi_{\rm E}$ as $\xi_{\rm E} < \xi_{\rm E, max}$ where $\xi_{\rm E, max}$ is defined as $\xi_{\rm E}$ of
Equation (\ref{eq-xiE}) with $M_C = M_{C, {\rm max}}$ \citep{ben08,sum10}. We applied the additional $\chi^2$ penalty presented by \citet{ben08} to 
each link of MCMC fitting;
\begin{align}
\chi^2_{\rm orb} = \Theta (\xi_{\rm E} - \xi_{\rm E, max}) \left( \frac{\xi_{\rm E} - \xi_{\rm E, max}}{\sigma_{\xi_{\rm E,max}}} \right)^2 \label{eq-chi2orb}
\end{align}
where $\Theta$ is the Heaviside step function and we applied 7$\%$ to $\sigma_{\xi_{\rm E,max}}$, the uncertainty of $\xi_{\rm E,max}$, 
with the consideration of uncertainty on $\theta_*$ as given in Section \ref{sec-color}.

We derive $D_S = 8.0 \pm 1.6$ kpc by a Bayesian analysis using the Galactic model \citep{han03} as the prior distribution constrained
 by the observed $t_{\rm E}$ value, 
$M_{S} = 1.02 \pm 0.12 M_{\odot}$ and $M_{C,{\rm max}} = 1.10 \pm 0.19 M_{\odot}$ from the color and brightness of the source and blending.
To calculate $\xi_{\rm E,max}$, we use the lower or upper limit value for each parameter so that it makes $\xi_{\rm E,max}$ larger for 
a conservative constraint, namely, we use $D_S = 6.4$ kpc, $M_{S} = 0.9 M_{\odot}$, $M_{C,{\rm max}} = 1.2 M_{\odot}$ for 
the calculation of $\xi_{\rm E,max}$. Note that we adopt $1.2 M_{\odot}$ as $M_{C,{\rm max}}$ considering that there are very few $M > 1.2 M_{\odot}$ stars in our galaxy \citep{gou00b, ben13}.
For $\rho$ and $P_{\xi}$, we used each link's values to calculate the $\xi_{\rm E,max}$.

The "xallarap" models in Table \ref{tab-models} are our results of xallarap fitting with a circular orbit and the $\xi_{\rm E,max}$ constraint above.
We find smaller $\chi^2$ values than that from the parallax models by $\Delta \chi^2 \gtrsim 27$.
Note that including eccentricity intends to fit systematics in the baseline and does not improve a model significantly,
so we don't consider eccentric orbits according to Occam's razor.

\subsection{Constraint on $\rho$ and lens properties}
Xallarap models place a lower limit on $\rho$ whereas the parallax models do not.
This is because the $\xi_{\rm E,max}$ constraint is equivalent to placing a lower limit of $\rho$ as
\begin{align}
\xi_{\rm E} &< \xi_{\rm E,max} \Leftrightarrow \rho > \rho_{\rm min} \equiv \frac{\theta_* D_S \xi_{\rm E}}{ M_{C, {\rm max}} (M_{C, {\rm max}}+M_S)^{-2/3} P_{\xi}^{2/3}}.
\end{align}
Combining it with the upper limit from the finite source effect, we can constrain the $\rho$ value with $\sim 30\%$ uncertainty.
This is the first case of $\rho$ being constrained with neither a significant finite source effect nor a parallax effect.
Then we can calculate a $\theta_{\rm E}$ value from $\rho$ and $\theta_*$,
\[ \theta_{\rm E} = 0.20 ^{+0.04}_{-0.08} \ {\rm mas}, \ \  \mu_{rel} = 1.1 ^{+0.2}_{-0.4} \ {\rm mas / yr} \]
for the xallarap close$+$ model.
Table \ref{tab-property} shows these values for all degenerate models using each models' values.

In principle, we can determine a mass and distance of the lens star by combining the $\theta_{\rm E}$ and 
$H_L$, the lens flux in $H$-band extracted from Keck AO observations \citep[e.g.]{bat14,fuk15}  because $\theta_{\rm E}$ gives us a mass-distance 
relationship;
\begin{align}
M_{\rm host} &= \frac{1}{1+q} \frac{\theta_{\rm E}^2}{\kappa \pi_{\rm rel}} \label{eq-mhdl3}
\end{align}
and we can convert the lens flux into another mass-distance relationship using a mass-luminosity relation.
However we encounter a problem with this.
As shown in Table \ref{tab-xallapro}, the uncertainty of the mass of the companion to the source star in the  xallarap models is very large and the upper limit of 
$\sim 1.8 M_{\odot}$ is larger than $M_{C,{\rm max}} = 1.2 M_{\odot}$, the maximum mass of the companion we use in the $\xi_{\rm E,max}$ constraint.
In other words, the lower limit of $H_C$, the apparent $H$ magnitude of the companion, is brighter than the brightness limit as
$H_C \simeq 16.2~ {\rm mag} < H_{\rm excess} \simeq 17.5~ {\rm mag}$ where $H_{\rm excess}$ value comes from Equation (\ref{eq-expara}).
It means we cannot place a fainter limit on $H_L$, the lens brightness in the $H$-band.

These excesses are attributed to the uncertainties of the parameters that determine $\xi_{\rm E, max}$, i.e., the uncertainties of $D_S$, $M_S$ and $\theta_*$.
Especially, it is more sensitive to the uncertainties of $\theta_*$ and $D_S$ rather than $M_S$ due to the relation between $M_C$ and the other parameters, 
$\xi_{\rm E} \propto (D_S \theta_*)^{-1} M_C (M_S + M_C)^{-2/3}$.
We calculate a $\xi_{\rm E,max}$ value with the most conservative combination of $D_S$ and $M_S$ in their 1 $\sigma$ uncertainties, 
$D_S=$ 6.4 kpc and $M_S= 0.9 M_{\odot}$, so that they make the $\xi_{\rm E,max}$ largest in the 1 $\sigma$ range.
Then we judge a set of ($\rho$, $P_{\xi}$, $\xi_{\rm E}$) by the conservatively large $\xi_{\rm E,max}$ value and accept them up to 
7$\%$ larger $\xi_{\rm E}$ than $\xi_{\rm E,max}$ considering the uncertainty on $\theta_*$.
Thus the $M_C$ value range derived from the accepted parameters with $D_S = 8.0 \pm 1.6$ kpc and $M_S = 1.02 \pm 0.12 M_{\odot}$ can exceed its limit,
$M_{C,{\rm max}} = 1.2 M_{\odot}$.

We can obtain only brighter limit of $H_L$ by subtracting the source brightness and the fainter limit of $H_C$ from $H_{\rm target}$ of Equation (\ref{eq-HKECK}),
\begin{equation}
H_L > H_{L,{\rm low}} = 17.66.  \label{eq-exxalla}
\end{equation}
Then we can place a lower limit of $M_{H,L}$ of $M_{H,L,{\rm low}} = H_{L,{\rm low}} -A_{H,L} - 5\log{(D_L/10 {\rm pc})}$ by using Equation (\ref{eq-mhdl1}) with $H_{L,{\rm low}}$.
Figure \ref{fig-MhDL-xalla} shows the mass-distance relation obtained from Equation (\ref{eq-mhdl3}) and $M_{H,L,{\rm low}}$.
From their overlapped region, we obtain constraints on the host mass and the distance as $M_{\rm host} < 1.16 M_{\odot}$ and $D_L < 9.3$kpc, respectively.
These constraints are very weak.

\subsection{Comparison with parallax model} \label{sec-compxp}
Xallarap models have $\chi^2$ differences from the parallax models by $\Delta \chi^2> 27$ for 3 dof difference.
\citet{poi05} analyzed 22 events where a parallax model improves their light curve fittings compared to the standard model.
According to their analysis, there are 3 events that prefer a xallarap model to a parallax model by $\Delta \chi^2> 25$ in all events they analyzed.
They regard it as a strong indication that the light curves of the 3 events have been distorted by xallarap.
{Here, we investigate whether the xallarap signal is real or unreal in the analyses bellow.}

We first plot $\chi^2$ values of the best-fit xallarap model at a fixed $P_{\xi}$ value within 
100 $\leq P_{\xi} ({\rm days}) \leq$ 1500 in Figure \ref{fig-chi2xalla}. They are shown as $\Delta \chi^2$ values compared to the parallax close$+$ model.
One of the standard ways to exclude a xallarap scenario is to show that just a narrow range of $P_{\xi}$ indicates a (small) preference in favor of the xallarap model over the parallax model, 
and then showing the very small probability that a binary system whose period is in such the narrow range happens to be microlensed \citep{ben08}.
For this event, however, the favored $P_{\xi}$ region compared to the parallax model is very broad even with the $\chi^2_{\rm orb}$ constraint.
Therefore, we cannot exclude the xallarap scenario by this approach.

\begin{figure}
\centering
\epsscale{0.6}
\rotatebox{-90}{
\plotone{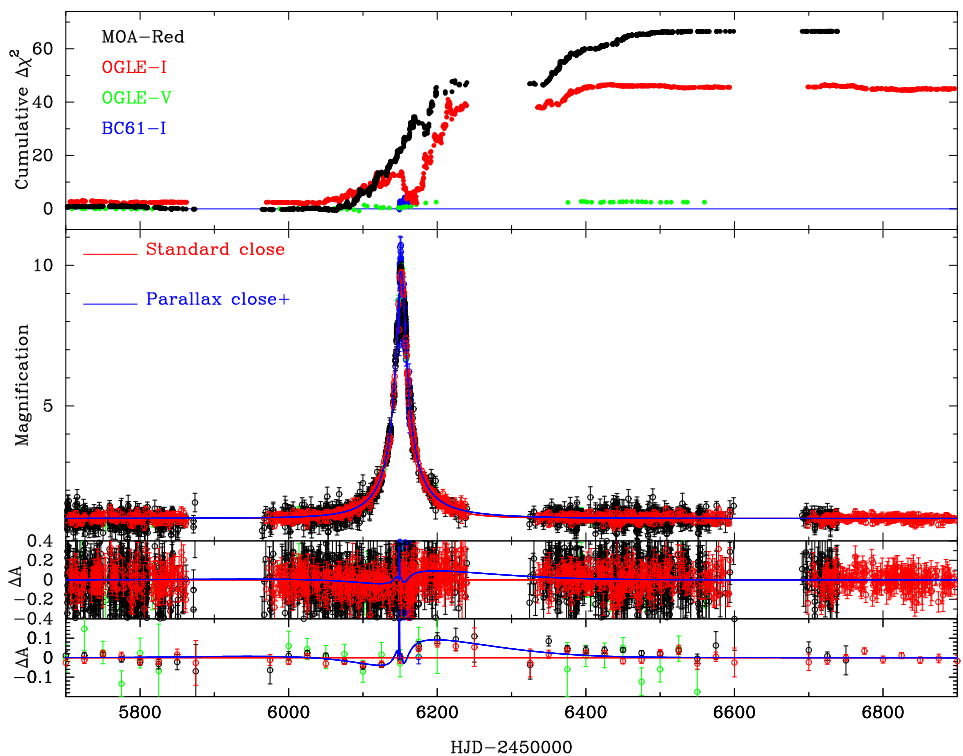}
}
\caption{Cumulative distribution of $\Delta \chi^2$ between the Standard close model and the parallax close$+$ model. Top inset shows the distribution and 
a positive $\Delta \chi^2$ value indicates smaller $\chi^2$ value of the parallax close$+$ than that of the Standard close model.
The second and third insets from the top shows the light curve and the residuals from the standard model, respectively. 
The bottom inset shows the residuals binned by 25 days for clarity.}
\label{fig-para-stand}
\end{figure}

\begin{figure}
\centering
\epsscale{0.7}
\rotatebox{-90}{
\plotone{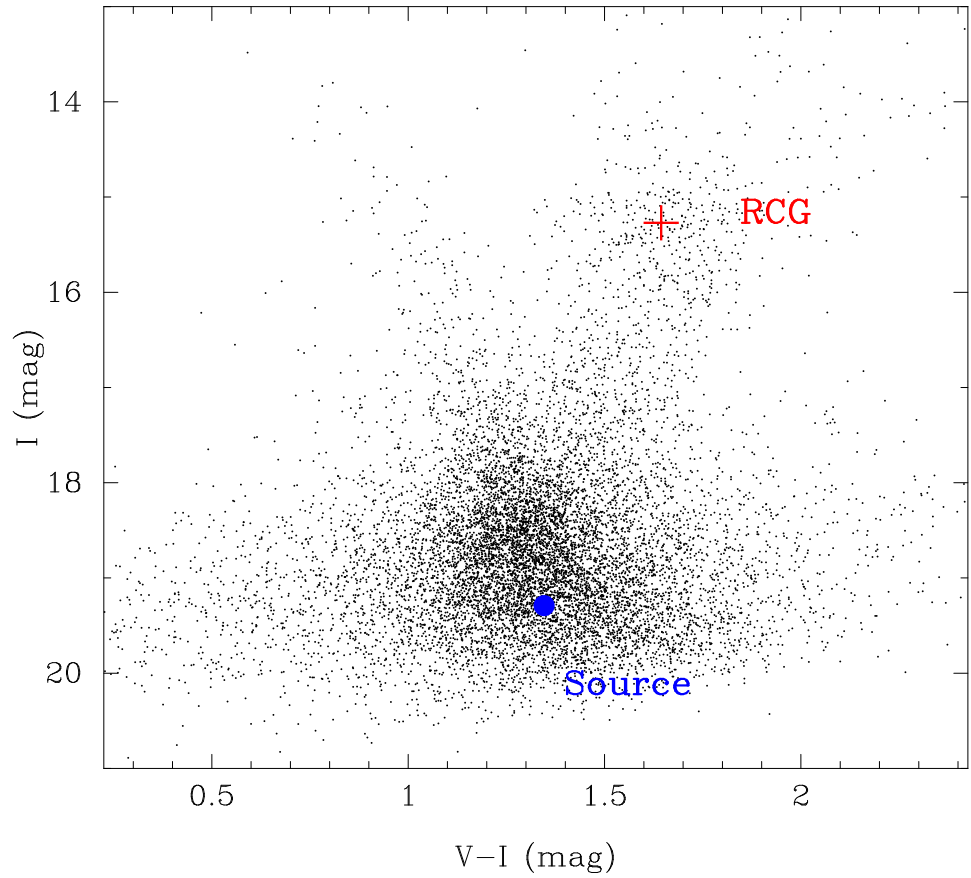}
}
\caption{The OGLE-IV instrumental color-magnitude diagram of stars within $2^{\prime} \times 2^{\prime}$ around the source star.
The source star and the mean of red clump giants are shown as the blue filled circle and the red cross, respectively.}
\label{fig-cmd}
\end{figure}

\begin{figure}
\begin{minipage}{0.5\hsize}
\begin{center}
\includegraphics[width=77mm]{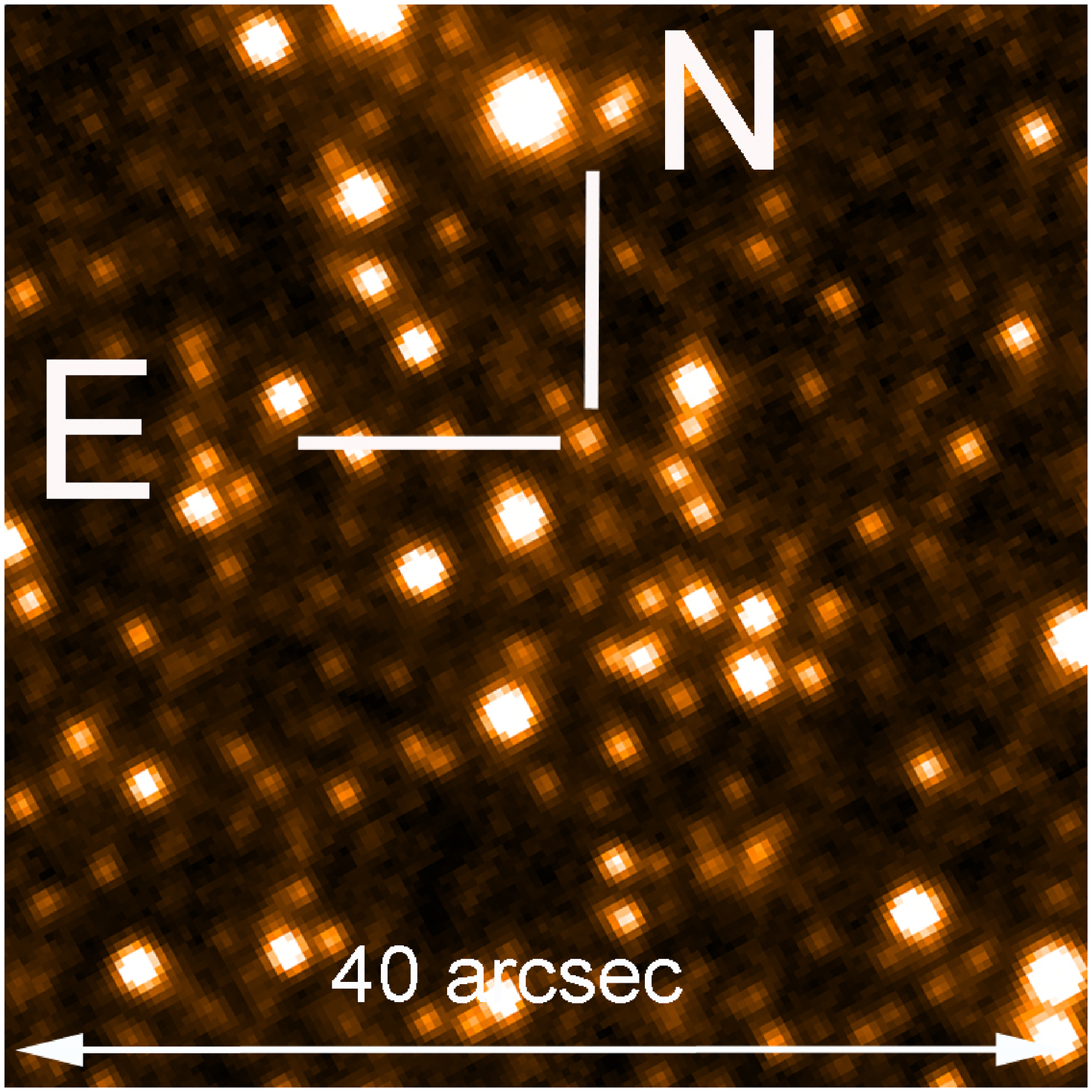}
\end{center}
\end{minipage}
\begin{minipage}{0.49\hsize}
\begin{center}
\includegraphics[width=77mm]{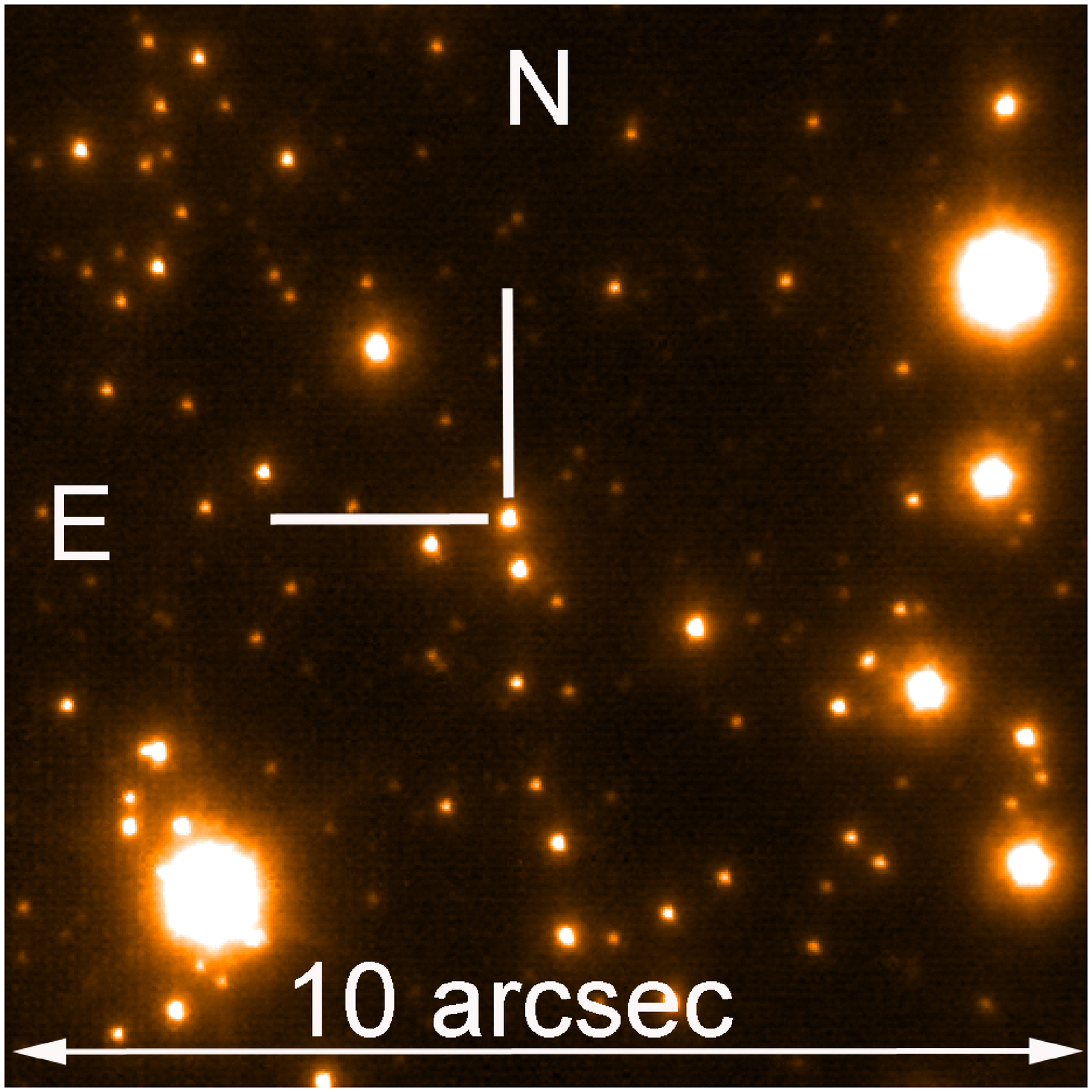}
\end{center}
\end{minipage}
\caption{Images of the event field observed by VVV (left) and by Keck II (right). The indicated star is the target.}
\label{fig-KECK}
\end{figure}

\begin{figure}
\centering
\epsscale{0.5}
\rotatebox{-90}{
\plotone{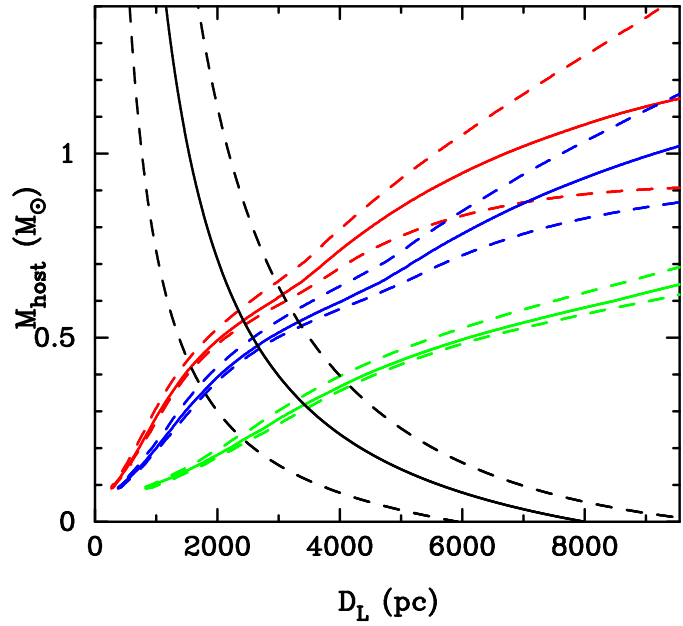}
}
\caption{Mass-distance relations for the parallax close$+$ model.
The relations from $M_{H,L}$ are shown in red, blue and green for a contamination fraction of $f=0$, 0.5 and 0.9, respectively.
The dashed lines indicate 1 $\sigma$ error including the uncertainty of the distance to the source, the lens age and the lens metallicity 
in addition to the uncertainty of our measuring.
Black lines are the mass-distance relation come from $\pi_{\rm E}$.
}
\label{fig-MhDL-para}
\end{figure}

\begin{figure}
\centering
\epsscale{0.7}
\rotatebox{-90}{
\plotone{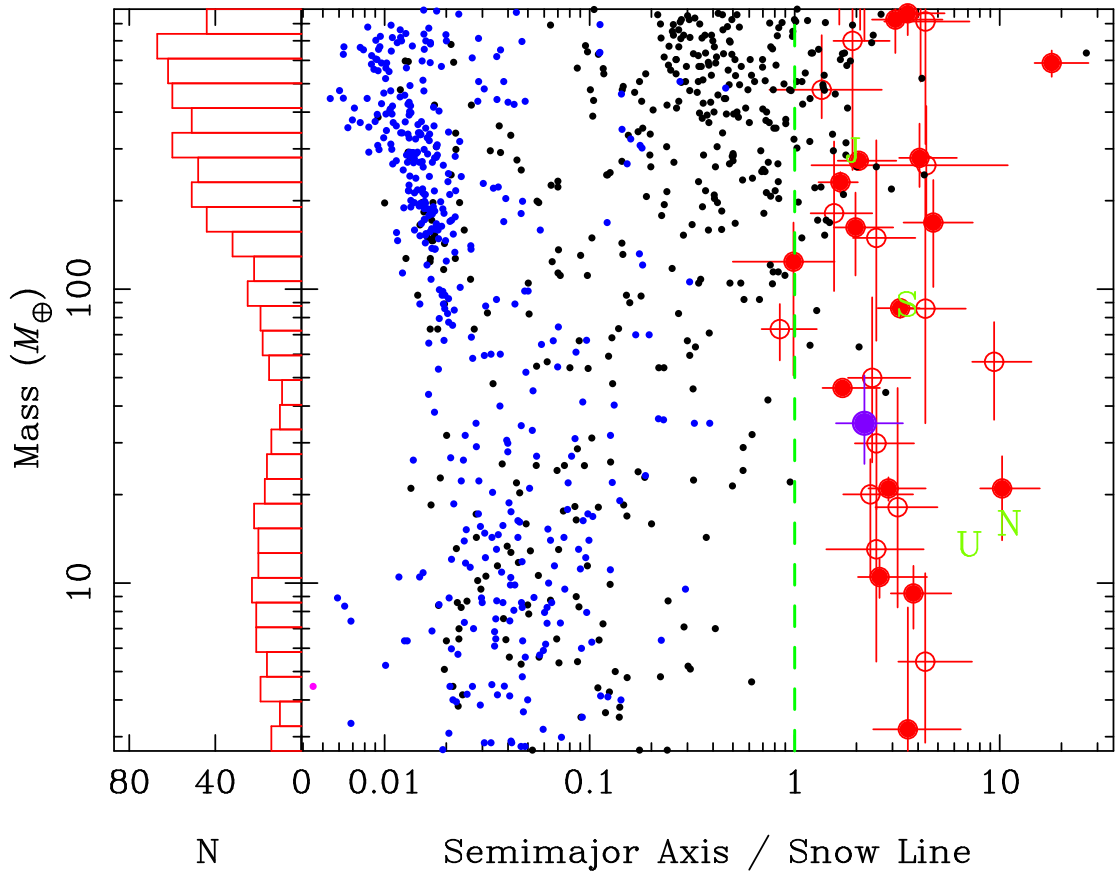}
}
\caption{Mass versus semi-major axis normalized by the snow-line of discovered exoplanets so far. 
Here the snow line is estimated by the host star mass as $\sim 2.7 M / M_{\odot}$ \citep{ken08}. 
The black and blue dots and red circles indicate planets found by the radial velocity, transit and microlensing, respectively. 
The results of this work are indicated as the purple circles. In microlensing planets, filled circles indicates that their masses are 
measured and open circles indicate that their masses are estimated by a Bayesian analysis. 
A green letter indicates a solar system planets.}
\label{fig-exo}
\end{figure}

\begin{figure}
\centering
\epsscale{0.5}
\plotone{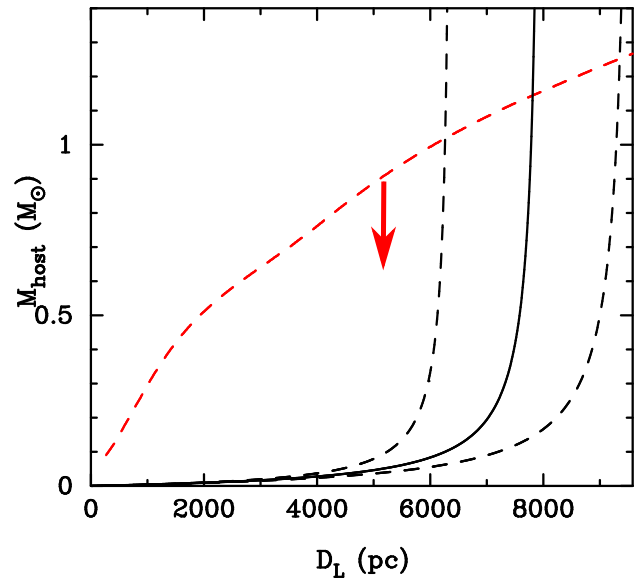}
\vspace{30pt}
\caption{Same as Figure \ref{fig-MhDL-para} but for the xallarap close$+$ model.
The relation from $M_{H,L, {\rm low}}$ (i.e., 1 $\sigma$ brighter limit of $M_{H,L}$ ) is shown in red.
The red arrow indicates that the red dashed line is just an upper limit.
Black lines are the mass-distance relation arising from $\theta_{\rm E}$ .
}
\label{fig-MhDL-xalla}
\end{figure}

\begin{figure}
\centering
\epsscale{0.5}
\rotatebox{-90}{
\plotone{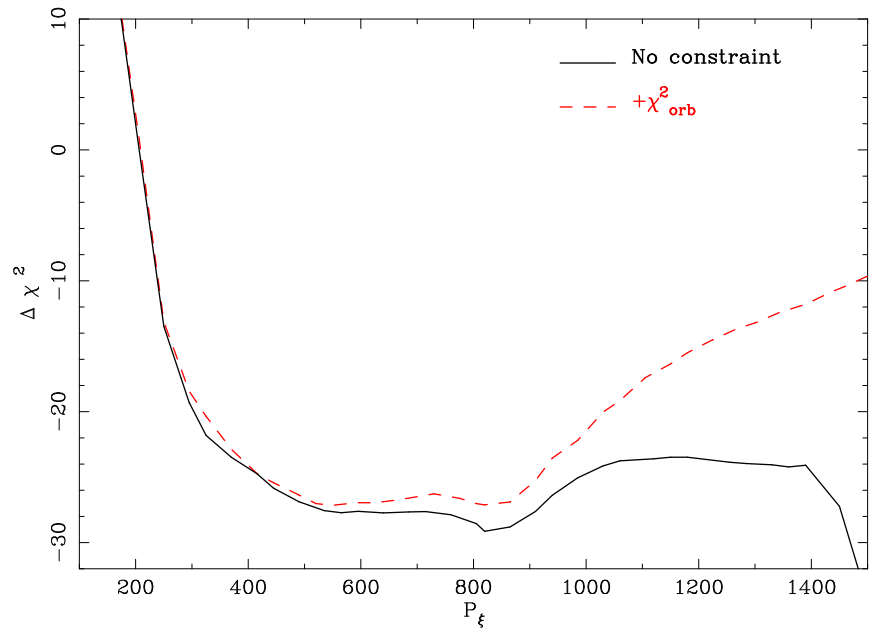}
}
\caption{$\Delta \chi^2$ between a xallarap model and the parallax close$+$ model as a function of orbital period.
The black solid and red dashed lines indicate xallarap models with no constraint and the $\chi^2_{\rm orb}$ constraint, respectively, see text.
}
\label{fig-chi2xalla}
\end{figure}

\begin{figure}
\centering
\epsscale{0.6}
\rotatebox{-90}{
\plotone{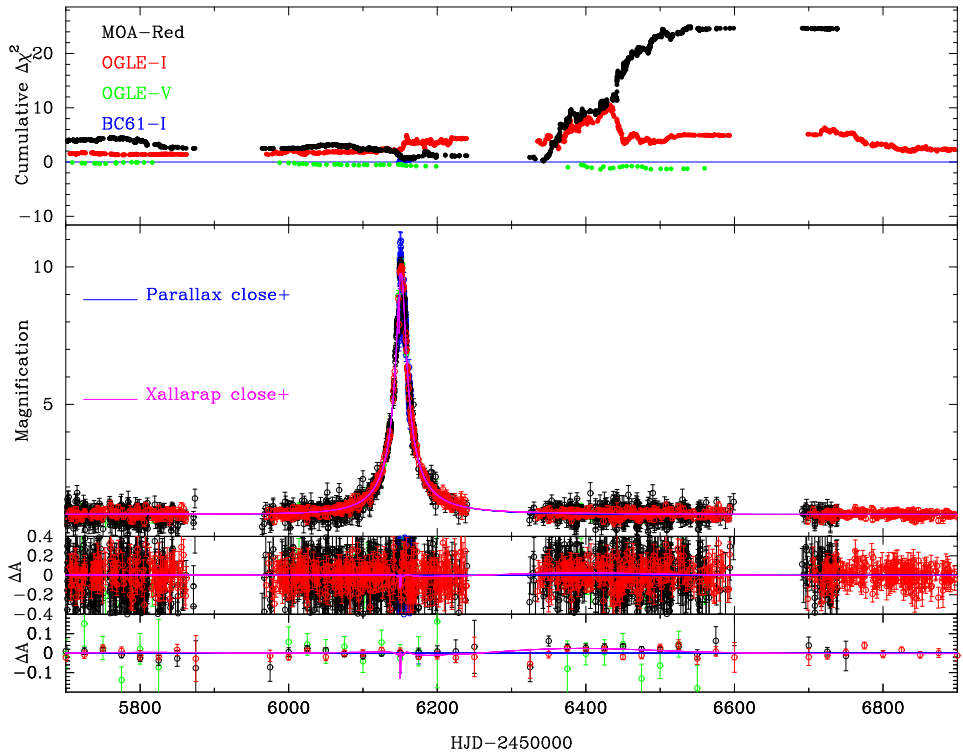}
}
\caption{As for Figure \ref{fig-para-stand}, but showing the difference between the parallax close$+$ model and the xallarap close$+$ model. 
A positive $\Delta \chi^2$ value indicates a smaller $\chi^2$ of the xallarap close$+$ model.}
\label{fig-xalla-para}
\end{figure}

Next, we investigate where the signal comes from in the light curve.
Figure \ref{fig-xalla-para} shows the same one as Figure \ref{fig-para-stand}, but shows the difference between the xallarap close$+$ model and 
the parallax close$+$ model.
The $\chi^2$ difference comes from the data in 2013 which is slightly magnified.
However, the preference to the parallax model comes almost exclusively from MOA data and the preference from OGLE data is only $\Delta \chi^2 \sim 2$.
The MOA data is more easily affected by systematics especially in a low-magnification part because the average seeing on the MOA site is worse than that on the OGLE site.
In addition to this inconsistency between MOA and OGLE, as shown in Table \ref{tab-xallapro}, a significant fraction of MCMC chains of every xallarap model indicates 
unphysically large masses of the companion even with the mass constraint imposed. These two facts are strong evidence against the xallarap model as the true model.
Thus we conclude that the signal modeled as xallarap in this model is unlikely to be owing to a real xallarap effect. Instead it is most likely to arise from low-level systematics in the MOA data.

Finally we fit the data consisting of only the anomaly part of MOA data (39 points) and entire OGLE $I$ data (1273 points) with both parallax and xallarap with no constraint  models
because the results are expected to be less affected by systematics.
Note that we use a potion of MOA data so that the best fit will not change largely except $\pi_{\rm E}$ or $\xi_{\rm E}$.
The xallarap model improves the fit by $\Delta \chi^2 = 6.5$ compared to the parallax model for 3 dof difference.
This indicates that the xallarap model is preferred by only $< 2\sigma$. Moreover, we compared these two models by the Bayesian information criterion (BIC), 
which is another common statistical criteria including penalty term for the number of fitting parameters, $BIC = \chi^2 + n_{\rm param} \ln (N_{\rm data})$ \citep{bur02}.
The $\Delta BIC$ is $BIC_{\rm para} - BIC_{\rm xalla} \sim -15$ and this criterion prefers the parallax model rather than the xallarap model.

Considering these facts, we conclude that the parallax model is preferred over the xallarap model.
Thus we dealt with only the parallax scenario in the main part of the paper.

\clearpage



\clearpage

\clearpage

\clearpage

\clearpage

\clearpage

\clearpage

\clearpage

\clearpage

\clearpage

\clearpage

\clearpage

\def\@captype{table}
 \tblcaption{The data and the parameters for our modeling.}
 \vspace{-0.4cm}
    \begin{center}
    \begin{tabular}{ccccc}\hline\hline
Dataset    & Number of data & $k$   & $e_{\rm min}$ & $u_{\lambda}$ \\\hline
OGLE $I$   & 1275           & 1.365 & 0             & 0.5470\\
OGLE $V$   &   81           & 1.576 & 0             & 0.7086\\
MOA-Red    & 6324           & 0.981 & 0             & 0.5895\\
B$\&$C $I$ &  382           & 1.017 & 0.00611       & 0.5470\\\hline
    \end{tabular}
    \end{center}
    \label{tab-data}
  \hfill

\clearpage

\begin{landscape}
{\tabcolsep = 1.2mm
\footnotesize
\catcode`?=\active \def?{\phantom{0}}
\caption{Model parameters.}
\vspace{-0.4cm}
\begin{center}
\begin{threeparttable}
    \begin{tabular}{lccccccccccccccccccccc}\hline\hline 
Model                 & $t_0$            & $t_{\rm E}$          & $u_0$				& $q$            & $s$                &$\alpha$            &$\rho$     		&$\pi_{E,N} / \xi_{E,N}$&$\pi_{E,E} / \xi_{E,E}$&$\pi_{\rm E} / \xi_{\rm E}$  &$P_{\xi}$       & $I_S$            &$\chi^2$&$dof$\\
                      &(HJD')            &(days)          &      				&($10^{-4}$)     &                    &(rad)               &($10^{-3}$)		&                      &                      &                   &($10^2$ days)   & (mag)            &        &     \\\hline
Standard        &           &            & 				&             &               &               &        		&                     &                     &                  &               &            &  & \\ 
????close        &6151.58           & 66.4           & 0.104				& 2.3            &0.890               & 5.098              &      4.3  		&       -              &       -              &    -              &      -         & 19.26            & 8168.8 &8047 \\ 
????$\sigma$              &   0.03           &  1.5           & 0.003				&$^{+0.6}_{-0.3}$&0.008               &$^{+0.009}_{-0.003}$&    $<7.0$ 		&       -              &       -              &    -              &      -         &  0.03            &   -    &  -  \\
????wide         &6151.59			 & 66.7           & 0.104				& 2.3            &1.007               & 5.100              &      3.4  		&       -              &       -              &    -              &      -         & 19.26            & 8168.1 &8047 \\ 
????$\sigma$              &   0.03			 &  1.6           & 0.003				&$^{+0.6}_{-0.2}$&0.008               &$^{+0.008}_{-0.004}$&    $<6.5$ 		&       -              &       -              &    -              &      -         &  0.03            &   -    &  -  \\\hline
Parallax        &           &            & 				&             &               &               &        		&                     &                     &                  &               &            &  & \\ 
????close$+$    &6151.47			 & 68.4           & 0.101				& 1.9            &0.895               & 5.081              &      0.0  		&     0.14             &     -0.19            &   0.24            &      -         & 19.29            & 8053.2 &8045 \\ 
????$\sigma$              &   0.03			 &  1.5           & 0.003				&$^{+0.5}_{-0.1}$&$^{+0.004}_{-0.007}$& 0.006              &    $<3.2$ 		&$^{+0.11}_{-0.06}$    &      0.02            &$^{+0.09}_{-0.04}$ &      -         &  0.03            &   -    &  -  \\
????close$-$    &6151.48			 & 68.5           &-0.100				& 1.9            &0.893               & 1.195              &      0.0  		&     0.08             &     -0.17            &   0.19            &      -         & 19.30            & 8055.8 &8045 \\ 
????$\sigma$              &   0.03			 &  1.6           & 0.003				&$^{+0.5}_{-0.2}$&0.006               &$^{+0.004}_{-0.007}$&    $<3.0$ 		&$^{+0.12}_{-0.07}$    &      0.02            &$^{+0.08}_{-0.02}$ &      -         &  0.03            &   -    &  -  \\ 
????wide$+$     &6151.49			 & 68.7           & 0.101				& 1.9            &1.004               & 5.083              &      0.0  		&     0.17             &     -0.20            &   0.26            &      -         & 19.29            & 8054.7 &8045 \\ 
????$\sigma$              &   0.03			 &  1.4           & 0.003				&$^{+0.6}_{-0.1}$&$^{+0.008}_{-0.004}$& 0.006              &    $<3.0$ 		&     0.08             &      0.02            &   0.06            &      -         &  0.03            &   -    &  -  \\
????wide$-$     &6151.48			 & 67.8           &-0.101				& 1.9            &1.004               & 1.195              &      0.0  		&     0.09             &     -0.17            &   0.19            &      -         & 19.29            & 8057.3 &8045 \\ 
????$\sigma$              &   0.02			 &  1.4           & 0.003               &$^{+0.6}_{-0.1}$&$^{+0.009}_{-0.003}$&$^{+0.003}_{-0.009}$&    $<3.1$ 		&$^{+0.10}_{-0.08}$    &      0.02            &$^{+0.07}_{-0.03}$ &      -         &  0.03            &   -    &  -  \\\hline
Xallarap      &           &            & 				&             &               &               &        		&                     &                     &                  &               &            &  & \\ 
????close$+$ &6151.50			 & 67.2           & 0.103               & 2.0            & 0.894              & 5.085              &      3.5       &     0.0              &     -0.6             &     0.6           &     5.3        & 19.27            & 8026.5 &8042 \\ 
????$\sigma$              &$^{+0.04}_{-0.02}$&$^{+2.8}_{-1.1}$&$^{+0.002}_{-0.005}$ &$^{+0.5}_{-0.2}$& 0.006              &$^{+0.012}_{-0.002}$&$^{+2.2}_{-0.6}$&$^{+0.1}_{-0.6}$      &$^{+1.0}_{-0.1}$      &$^{+0.3}_{-0.2}$   &$^{+1.9}_{-0.4}$&$^{+0.06}_{-0.02}$&   -    &  -  \\ 
????close$-$ &6151.50			 & 66.8           &-0.103               & 2.0            & 0.895              & 1.197              &      3.8       &     0.1              &     -0.6       	  &     0.6           &     5.3        & 19.26            & 8026.4 &8042 \\ 
????$\sigma$              &$^{+0.04}_{-0.01}$&$^{+3.6}_{-0.7}$&$^{+0.006}_{-0.001}$ &$^{+0.5}_{-0.2}$& 0.006              &$^{+0.003}_{-0.010}$&$^{+2.0}_{-0.8}$&$^{+0.1}_{-0.6}$      &$^{+1.0}_{-0.2}$	  &$^{+0.4}_{-0.2}$   &$^{+2.4}_{-0.5}$&$^{+0.07}_{-0.02}$&   -    &  -  \\
????wide$+$  &6151.50			 & 66.7           & 0.104               & 2.1            & 1.005              & 5.088              &      2.9       &    -0.5              &     -0.3       	  &     0.6           &     5.2        & 19.26            & 8027.5 &8042 \\ 
????$\sigma$              &$^{+0.04}_{-0.01}$&$^{+3.6}_{-0.9}$&$^{+0.002}_{-0.006}$ &$^{+0.5}_{-0.3}$&$^{+0.007}_{-0.006}$&$^{+0.009}_{-0.005}$&$^{+2.6}_{-0.1}$&$^{+0.6}_{-0.1}$      &$^{+0.1}_{-0.6}$	  &$^{+0.5}_{-0.2}$   &$^{+2.7}_{-0.4}$&$^{+0.07}_{-0.02}$&   -    &  -  \\
????wide$-$  &6151.50			 & 66.4           &-0.104               & 2.3            & 1.006              & 1.194              &      3.6       &     0.0              &     -0.6       	  &     0.6           &     5.3        & 19.26            & 8027.9 &8042 \\ 
????$\sigma$              &$^{+0.04}_{-0.01}$&$^{+4.0}_{-0.3}$&$^{+0.007}_{-0.001}$ &$^{+0.3}_{-0.4}$&$^{+0.006}_{-0.007}$&$^{+0.006}_{-0.009}$&$^{+2.0}_{-0.7}$&$^{+0.1}_{-0.6}$      &$^{+1.1}_{-0.2}$	  &$^{+0.5}_{-0.2}$   &$^{+2.5}_{-0.5}$&$^{+0.08}_{-0.01}$&   -    &  -  \\\hline
    \end{tabular}
    \begin{tablenotes}
    \item { A superscript/subscript indicates the difference parameter's 84/16 percentile from the best-fit value, respectively.}
    \end{tablenotes}
    \end{threeparttable}
    \end{center}
    \label{tab-models}
}
\end{landscape}

\normalsize

\clearpage

{\tabcolsep = 1.2mm
\scriptsize
\catcode`?=\active \def?{\phantom{0}}
\caption{Lens properties.}
\vspace{-0.6cm}
\begin{center}
\begin{threeparttable}
    \begin{tabular}{lccccccccccccccccccccc}\hline\hline 
Model                     &$P$($f' \leq f$) \tnote{1}& $\theta_*$& $\theta_{\rm E}$   & $\mu_{rel}$       & $M_{\rm host}$    	  & $M_{\rm p}$    & $D_{L}$           & $r_{\perp}$       &$a$ \\
					      &   $\%$    & ($\mu$as)     	 &   (mas) 		        &   (mas/yr)        &($M_{\odot}$)   	  &($M_{\oplus}$)  & (kpc)             & (AU)              &  (AU)             \\\hline
Parallax                  &           &               	 &                      &                   &                	  &                &                   &                   &                   \\ 
????close$+$              &    -      &0.69 $\pm$ 0.05	 &  $>0.22$		        &   $>1.2$          &    $>0.10$     	  &     $>7$       &   $<5.6$          &     -             &     -             \\ 
 ????? w/ KECK ($f$=0)    &   62.6     &      -        	 &1.09$^{+0.16}_{-0.10}$&5.8$^{+0.8}_{-0.6}$&0.57$^{+0.06}_{-0.10}$&35$^{+10}_{-6}$&2.6$^{+0.5}_{-0.7}$&2.6$^{+0.3}_{-0.5}$&3.1$^{+1.5}_{-0.7}$\\
 ????? w/ KECK ($f$=0.5)  &  89.3    &      -        	 &0.96$^{+0.12}_{-0.09}$&5.1$^{+0.6}_{-0.5}$&0.50$^{+0.05}_{-0.09}$&31$^{+9}_{-5}$ &2.9$^{+0.6}_{-0.8}$&2.4$^{+0.2}_{-0.4}$&3.0$^{+1.4}_{-0.5}$\\
 ????? w/ KECK ($f$=0.9)  &  95.4    &      -        	 &0.64$^{+0.06}_{-0.05}$&3.4$ \pm 0.3$      &0.33$^{+0.05}_{-0.08}$&21$^{+6}_{-4}$ &3.6$^{+0.6}_{-0.8}$&2.1$^{+0.2}_{-0.3}$&2.5$^{+1.2}_{-0.6}$\\
????close$-$              &    -      &0.69 $\pm$ 0.05	 &  $>0.22$             &   $>1.2$          &    $>0.13$     	   &    $>9$       &   $<5.8$          &     -             &     -             \\ 
 ????? w/ KECK ($f$=0)    &   61.9     &      -      	 &0.99$^{+0.18}_{-0.06}$&5.3$^{+1.0}_{-0.3}$&0.63$^{+0.04}_{-0.11}$&41$^{+9}_{-7}$ &3.2$^{+0.4}_{-0.9}$&2.8$^{+0.2}_{-0.5}$&3.4$^{+1.6}_{-0.8}$\\ 
 ????? w/ KECK ($f$=0.5)  &  88.8     &      -        	 &0.86$^{+0.15}_{-0.06}$&4.6$^{+0.9}_{-0.3}$&0.55$^{+0.03}_{-0.09}$&36$^{+8}_{-6}$ &3.4$^{+0.4}_{-0.9}$&2.7$^{+0.1}_{-0.4}$&3.3$^{+1.5}_{-0.7}$\\
 ????? w/ KECK ($f$=0.9)  &  95.1     &      -        	 &0.60$^{+0.08}_{-0.04}$&3.2$^{+0.4}_{-0.2}$&0.38$^{+0.03}_{-0.08}$&25$\pm 5$      &4.2$^{+0.4}_{-0.9}$&2.2$^{+0.1}_{-0.3}$&2.7$^{+1.3}_{-0.6}$\\
????wide$+$               &    -      &0.69 $\pm$ 0.05	 &  $>0.23$             &   $>1.2$          &    $>0.11$       	   &    $>8$       &   $<5.5$          &     -             &     -             \\ 
 ????? w/ KECK ($f$=0)    &   63.3     &      -     		 &1.14$^{+0.11}_{-0.15}$&6.1$^{+0.6}_{-0.8}$&0.54$^{+0.09}_{-0.07}$&34$^{+13}_{-4}$&2.4$^{+0.8}_{-0.5}$&2.7$\pm 0.4$       &3.3$^{+1.9}_{-0.7}$\\
 ????? w/ KECK ($f$=0.5)  &  89.4    &      -        	 &0.99$^{+0.08}_{-0.13}$&5.3$^{+0.4}_{-0.7}$&0.47$^{+0.08}_{-0.07}$&30$^{+11}_{-4}$&2.6$^{+0.8}_{-0.5}$&2.6$\pm 0.4$       &3.2$^{+1.8}_{-0.6}$\\
 ????? w/ KECK ($f$=0.9)  &  95.6    &      -        	 &0.66$^{+0.04}_{-0.07}$&3.5$^{+0.2}_{-0.4}$&0.31$^{+0.07}_{-0.05}$&20$^{+8}_{-3}$ &3.4$^{+0.8}_{-0.5}$&2.2$\pm 0.3$       &2.7$^{+1.5}_{-0.5}$\\
????wide$-$               &    -      &0.69 $\pm$ 0.05	 &  $>0.23$             &   $>1.2$          &    $>0.13$       	   &    $>9$       &   $<5.8$          &     -             &     -             \\ 
 ????? w/ KECK ($f$=0)    &  62.1     &      -    		 &0.99$^{+0.17}_{-0.06}$&5.3$^{+0.9}_{-0.4}$&0.63$^{+0.05}_{-0.10}$&39$^{+13}_{-4}$&3.2$^{+0.4}_{-0.9}$&3.1$^{+0.2}_{-0.5}$&3.8$^{+1.9}_{-0.8}$\\ 
 ????? w/ KECK ($f$=0.5)  &  88.9     &      -        	 &0.87$^{+0.14}_{-0.07}$&4.7$^{+0.8}_{-0.4}$&0.55$^{+0.04}_{-0.09}$&34$^{+11}_{-4}$&3.4$^{+0.4}_{-0.9}$&3.0$^{+0.2}_{-0.4}$&3.6$^{+1.8}_{-0.8}$\\
 ????? w/ KECK ($f$=0.9)  &  95.1     &      -        	 &0.60$^{+0.07}_{-0.04}$&3.2$^{+0.4}_{-0.2}$&0.38$^{+0.04}_{-0.08}$&24$^{+8}_{-3}$ &4.2$^{+0.4}_{-0.9}$&2.5$^{+0.1}_{-0.3}$&3.1$^{+1.5}_{-0.6}$\\
????{\bf Mean}          &    -      &{\bf 0.69 $\pm$ 0.05}	 &{\bf 0.99}$^{\bf +0.26}_{\bf -0.19}$&{\bf 5.3}$^{\bf +1.4}_{\bf -1.0}$&{\bf 0.56}$^{\bf +0.12}_{\bf -0.16}$&{\bf 35}$^{\bf +17}_{\bf -9}$&{\bf 3.0}$^{\bf +0.8}_{\bf -1.1}$&{\bf 2.7}$^{\bf +0.6}_{\bf -0.7}$&{\bf 3.4}$^{\bf +2.4}_{\bf -1.1}$\\\hline
 Xallarap      &       &                 	 &         		        &                   &                	   &               &                   &                   &                   \\ 
????close$+$              &   -   &0.70$^{+0.05}_{-0.06}$&0.20$^{+0.04}_{-0.08}$&1.1$^{+0.2}_{-0.4}$&      -         	   &     -         &   -               &     -             &     -    \\ 
 ????? w/ KECK            &   -   &      -        		 &        -             &			-		& $< 1.16$             &   $<77$       &  $<9.3$           &    $<1.7$         &     -             \\ 
????close$-$              &   -   &0.70$^{+0.05}_{-0.06}$&0.19$^{+0.05}_{-0.07}$&1.0$^{+0.3}_{-0.4}$&      -      	       &     -         &   -               &     -             &     -             \\ 
????wide$+$               &   -   &0.70$^{+0.05}_{-0.06}$&0.24$^{+0.00}_{-0.11}$&1.3$^{+0.0}_{-0.6}$&      -               &     -         &   -               &     -             &     -             \\ 
????wide$-$               &   -   &0.70$^{+0.04}_{-0.07}$&0.20$^{+0.05}_{-0.07}$&1.1$^{+0.2}_{-0.4}$&      -               &     -         &   -               &     -             &     -             \\\hline 
    \end{tabular}
    \begin{tablenotes}
    \item A superscript or subscript of $+$/$-$ 0.0 indicates that the parameter's best fit value is same as the 84/16 percentile to the given significant digits
    \footnotesize
    \item[1] Probability of contamination fraction not exceeding $f$.
    \end{tablenotes}
    \end{threeparttable}
    \end{center}
\label{tab-property}
}

\clearpage

\small
\catcode`?=\active \def?{\phantom{0}}
\caption{1 $\sigma$ possible value ranges of mass and $H$ magnitude of the source companion for xallarap models.}
\vspace{-0.4cm}
\begin{center}
\begin{threeparttable}
    \begin{tabular}{lccccccccccccccccccccc}\hline\hline 
Model                 &   $M_C$        &   $H_C$  	   \\       
					  & ($M_{\odot}$)  &   (mag) 	        \\\hline 
Xallarap        &                &               &                     \\       
????close$+$          &   0.57 - 1.71  &  16.3 - 20.4          \\       
????close$-$          &   0.56 - 1.71  &  16.3 - 20.5        \\       
????wide$+$           &   0.62 - 1.81  &  16.1 - 20.7         \\       
????wide$-$           &   0.60 - 1.76  &  16.2 - 20.2         \\\hline      
    \end{tabular}
    \begin{tablenotes}
    \item The range of 1 $\sigma$ error is shown for each parameter.
    \end{tablenotes}
    \end{threeparttable}
    \end{center}
\label{tab-xallapro}

\end{document}